Factor endowment–commodity output relationships in a three-factor, two-good general equilibrium trade model

Yoshiaki Nakada

Division of Natural Resource Economics, Graduate School of Agriculture, Kyoto University, .Sakyo, Kyoto, Japan.

E-mail: nakada@kais.kyoto-u.ac.jp

Abstract:

We analyze the Rybczynski sign pattern, which expresses the factor endowment–commodity output relationships in a three-factor, two-good general equilibrium trade model. The relationship determines whether a strong Rybczynski result holds. We search for a sufficient condition for each Rybczynski sign pattern to hold in a systematic manner, which no other studies have derived. We assume factor-intensity ranking is constant. We use the EWS (economy-wide substitution)-ratio vector and the Hadamard product in our analysis. We show that the position of the EWS-ratio vector determines the Rybczynski sign pattern. This article provides a basis for further applications.

JEL D50, D58, F10, F11

Acknowledgement:

We would like to thank Editage (www.editage.jp) for English language editing. I am grateful to Dr. Teramachi Nobuo, for reading an earlier version of this article and an earlier version of Appendix A (Nakada 2015b).

Equation Section 1

1. Introduction

Batra and Casas (1976) (hereinafter BC) published an article on functional relations in a three-factor, two-good neoclassical model (or 3 x 2 model). The authors claimed that 'a strong Rybczynski result' arises if we use Thompson's (1985) terminology. According to Suzuki (1983, p. 141), BC contended in Theorem 6 (p. 34) that 'if commodity 1 is relatively capital intensive and commodity 2 is relatively labor intensive, an increase in the supply of labor increases the output of commodity 2 and reduces the output of commodity 1. [Moreover, an increase in the supply of capital increases the output of commodity 1 and reduces the output of commodity 2.]'[1] This is what a strong Rybczynski result implies. A strong Rybczynski result is a loose concept, as we show later, and it includes three Rybczynski sign patterns, which express the factor endowment–commodity output relationships.

Suzuki (1983) contended that this could not be the case under the assumption of 'perfect complementarity.' He used the Allen partial elasticities of substitution (hereinafter AES) for his analysis. Jones and Easton (1983) (hereinafter JE) mainly analyzed the commodity price–factor price relationship. This relationship is the dual counterpart in the factor endowment–commodity output relationship. On this duality, see JE (p. 67); see also BC (p. 36, eqs. (31)-(33)). In section 4 (pp. 77-81), JE showed six patterns of the commodity price–factor price relationship using a diagrammatic technique.[2] Apparently, a strong Rybczynski result does not hold for some relationships that JE showed. JE (p. 75) defined 'economy-wide substitution' (hereinafter EWS) for their analysis. Using EWS, JE showed some sufficient conditions for the commodity price–factor price relationship to hold in subsection 5.2 (pp. 86-92). JE suggested that 'the factor-intensity ranking' and EWS are important for their analysis (see p. 67 and 96). Thompson (1985) also tried to show some sufficient conditions for a strong Rybczynski result to hold (or not to hold). He used the concept of 'aggregate substitution.' Aggregate substitution is related with EWS as shown in eq. (A16).

In summary, these three articles tried to disprove BC's claim of 'a strong Rybczynski result' and tried to show sufficient conditions for that result to hold (or not to hold). However, Suzuki (1983)'s proof is not plausible (see Nakada (2015a)).[3] JE's analysis is somewhat complicated. Particularly, JE's proof in subsections 5.2.4 and 5.2.5 (p. 90-2) using 'perfect complementarity', defined by the authors themselves, is implausible (see Nakada (2015b) and eq. (A18)).[4] Thompson's analysis is questionable. In the Appendix (p. 66-70), Teramachi (1993) commented that the analysis in Thompson (1985) was not plausible. Before Thompson (1985), it was meaningful to disprove the results derived by BC; however, since Thompson (1985), the significance of disproving the results seems to have decreased.

On the other hand, Takayama (1982, p. 13-21) analyzed the factor endowment–commodity output relationship and its dual counterpart in the 3 x 2 model in his survey article. According to Takayama, if 'extreme factors' are 'aggregate complements' (for this definition, see Takayama, (1982, p. 18)), we derive the result that is equivalent to 'a strong Rybczynski result.'[5]

The following questions arise.

(i)     Can we do a more detailed analysis on a sufficient condition for each Rybczynski sign pattern to hold? If so, how can we do it?

(ii)    What results can we derive otherwise?

After Thompson (1985), what studies have been conducted on the 3 x 2 model? We explain the articles that address the factor endowment–commodity output relationship and/or its dual-counterpart, the commodity price–factor price relationship.

We have classified the articles after Thompson (1985) as follows.[6]

(i) Studies that assume the functional form of production functions; for example, Thompson (1995).

(ii) Studies that make another assumption concerning production functions (e.g., normal property, separability). See, for example, Suzuki (1985), Suzuki (1987, Chapter 2), and Bliss (2003).

(iii) Studies that modify one of the basic assumptions; for example, Ide (2009).

(iv) Other studies, for example, Teramachi (1993, 1995, 2015) and Easton (2008).

(v) Studies that analyzed a somewhat different aspect or the commodity price–relative factor price relationship. For example, Ban (2007a), Ban (2008), and Ban (2011). See also Ban (2007b). Ban assumed that production functions are of the two-level CES type.

In summary, some of these studies after Thompson (1985) are more complex. I am uncertain whether all of these studies have plausible results. Some papers apply the models before the basic functions of the model are understood. The analyses in some articles differ significantly from others, and it is not easy to make a comparison. Some studies such as Bliss (2003) did not present the process of computation. Some results are implausible. For example, Easton (2008) tried to extend the concept of 'perfect complementarity' defined by JE, which does not hold, as the author showed in the Appendix A (see eq. (A18)). Therefore, it seems questionable to extend that concept further as in Easton (2008). Teramachi (1993, 2015) assumed that, for example, extreme factors are perfect complements, as JE assumed. However, that is implausible, as I stated earlier.[7]

At least, concerning a sufficient condition for each Rybczynski sign pattern to hold in the 3 x 2 model of BC's original type, other studies are far from systematic. That is, no other studies derived all the conditions in a one-to-one correspondence. The purpose of this article is to derive such a condition in a systematic manner. Notably, we define the EWS-ratio vector based on the EWS,[8] and we use this and the Hadamard product of matrices for the analysis. The EWS-ratio is the relative magnitude of EWS compared to another EWS. In this article, we conclude that the position of the EWS-ratio vector determines the Rybczynski sign pattern. Using this relationship, we derive a sufficient condition for a strong Rybczynski result to hold (or not to hold).

Can we estimate the position of the EWS-ratio vector? Nakada (2016a) proves that we can estimate it if we have appropriate data. This article provides a basis for further applications. For example, it is useful for estimating the Rybczynski sign pattern in some countries, and it will contribute to international economics and energy economics.[9] Section 2 of this study explains the model. In subsection 2.1, we explain the basic structure of the model. We make a system of linear equations using a 5 x 5 matrix.[10] In subsection 2.2, we assume factor-intensity ranking.[11] In subsection 2.3, we define the EWS-ratio vector based on EWS for the analysis. We derive the important relationship among EWS-ratios and draw the EWS-ratio vector boundary in the figure, which is useful for our analysis. In subsection 2.4, we derive the solutions of a system of linear equations. In subsection 2.5, we develop a Rybczynski matrix and transform its component using EWS-ratios. In subsection 2.6, we draw the border line for a Rybczynski sign pattern to change in the figure, which we call line *ij*. Line *ij* divides the region of the EWS-ratio vector into 12 subregions. In subsection 2.7, we analyze Rybczynski sign patterns using the Hadamard product of matrices and derive a sufficient condition for each Rybczynski sign pattern to hold and, next, derive a sufficient condition for a strong Rybczynski result to hold (or not to hold). In subsection 2.8, we analyze Stolper-Samuelson sign patterns, which express the commodity price–factor price relationships. In section 3, we show some applications of these results. Section 4 presents the conclusion, and the Appendix A derives the important relationship among EWSs. The Appendix B shows that the determinant of the coefficient matrix of a system of linear equations, is negative.

The studies after Thompson (1985) are as follows.

(i) Thompson (1995) assumed that production functions were of the trans-log type, and estimated the values of parameters in the United States using econometrics. Based on these parameter values, he computed 'the aggregate elasticities' (equivalent to EWS). This is an application.[12] Next, Thompson assumed that production functions were of the Cobb-Douglas and CES types. Additionally, he assumed 'strong degrees of complementarity.' This is a simulation.

(ii) Suzuki (1985) assumed 'normal property' of the factor of production. In Suzuki (1987, Chapter 2, p. 27-36), the

author assumed that production functions were separable (p. 32). Bliss (2003) assumed that only one sector had a specific factor. He assumed separability and non-separability in production functions. Bliss (2003) assumed that capital and land were 'Hicksian complements' in agriculture (p. 274) and attempted to explain the wage movement in British economic history. This is a type of application.

(iii) Ide (2009) modified one of the model's basic assumptions and assumed the model with increasing returns to scale technology. He assumed that extreme factors were 'aggregate complements.' This is a theoretical study.

(iv) Teramachi (1993) analyzed the commodity price–factor price relationships in elasticity terms. Teramachi (2015, Chapter 3) published most of this article as a chapter in his book.[13] See also Teramachi (1995). Easton (2008) analyzed whether the extent of substitutability and complementarity affected the commodity price–factor price relationships. He reconsidered the analysis in JE (1983).

(v) Ban (2007a) attempted to analyze how commodity prices affected relative factor prices, for example, the skilled labor wage over unskilled labor wage. The author described the effects when she changed 'the cost share pattern' (or the factor-intensity ranking in our expression). She assumed that production functions were of the two-level CES type. In her model, the three factors are skilled labor, capital, and unskilled labor. She assumed that skilled labor and capital could be '[Allen-] complements' in each sector, and she computed the values of AES theoretically. However, her analysis is somewhat complicated, and her results are not clear. She showed some definite results. This is a theoretical study. Ban (2008, p. 4, Table 1) showed a table classifying the results in Ban (2007a) by factor-intensity ranking and factor-intensity ranking for the middle factor.[14] She classified the countries in the world into 14 regions in total and computed the input-output coefficient for each area using the GTAP version 6 database to derive factor-intensity ranking. However, Ban (2008) did not show the factor-intensity ranking for the middle factor.[15] Additionally, she assumed 10 types of values for 'the elasticities of substitution' (equivalent to EWS) to simulate how commodity prices affect the relative factor prices. This is an application. Ban (2011, chapter 4, p. 87-109) summarized the results of Ban (2007a) and Ban (2008) and modified the studies. For her results, see Ban (2011, p. 96-7, Table 4-1). See also Ban (2007b). In summary, in all her studies, Ban assumed that production functions are of the two-level CES type. Therefore, I am uncertain whether the results hold in general.

2. Model

2.1. Basic structure of the model

We assume similarly to BC (p. 22-3). That is, we assume as follows. Products and factor markets are perfectly competitive. The supply of all factors is perfectly inelastic. Production functions are homogeneous of degree one and strictly quasi-concave. All factors are not specific and perfectly mobile between sectors, and factor prices are perfectly flexible. These two assumptions ensure the full employment of all resources. The country is small and faces exogenously given world prices, or the movement in the price of a commodity is exogenously determined. The movements in factor endowments are exogenously determined.

Full employment of factors implies

$$\sum_j a_{ij} X_j = V_i, \quad i = T, K, L \tag{1}$$

where $X_j$ denotes the amount produced of good $j$ ($j = 1, 2$); $a_{ij}$ denotes the requirement of input $i$ per unit of output of good $j$ (or the input-output coefficient); $V_i$ denotes the supply of factor $i$; $T$ is the land, $K$ capital, and $L$ labor.

In a perfectly competitive economy, the unit cost of production of each good must just equal its price. Hence,

$$\sum_i a_{ij} w_i = p_j, \quad j = 1, 2 \tag{2}$$

where $p_j$ is the price of good $j$, and $w_i$ is the reward of factor $i$.

BC (p. 23) stated, 'With quasi-concave and linearly homogeneous production functions, each input-output coefficient is independent of the scale of output and is a function solely of input prices:'

$$a_{ij} = a_{ij}(w_i), \quad i = T, K, L, \quad j = 1, 2. \tag{3}$$

The authors continued, 'In particular, each $C_{ij}$ [$a_{ij}$ in our expression] is homogeneous of degree zero in all input prices.'[16]

Equations (1)-(3) describe the production side of the model. These are equivalent to eqs (1)-(5) in BC. The set includes 11 equations in 11 endogenous variables ($X_j$, $a_{ij}$, and $w_i$) and five exogenous variables ($V_i$ and $p_j$). The small country assumption simplifies the demand side of the economy. Totally differentiate (1):

$$\Sigma_j (\lambda_{ij} a_{ij}^* + \lambda_{ij} X_j^*) = V_i^*, \quad i = T, K, L, \tag{4}$$

where an asterisk denotes a rate of change (e.g., $X_j^* = dX_j / X_j$), and where $\lambda_{ij}$ is the proportion of the total supply of factor $i$ in sector $j$ (that is, $\lambda_{ij} = a_{ij} X_j / V_i$). Note that $\Sigma_j \lambda_{ij} = 1$.

The minimum unit cost equilibrium condition in each sector implies $\Sigma_i w_i da_{ij} = 0$. Hence, we derive (see JE (p. 73), BC (p. 24, note 5),

$$\Sigma_i \theta_{ij} a_{ij}^* = 0, \ j = 1, \ 2, \tag{5}$$

where $\theta_{ij}$ is the distributive share of factor $i$ in sector $j$ (that is, $\theta_{ij} = a_{ij} w_i / p_j$). Note that $\Sigma_i \theta_{ij} = 1$; $da_{ij}$ is the differential of $a_{ij}$.

Totally differentiate eq. (2):

$$\Sigma_i \theta_{ij} w_i^* = p_j^*, \ j = 1, \ 2. \tag{6}$$

Subtract $p_1^*$ from both sides of eq. (6):

$$\Sigma_i \theta_{i1} w_{i1}^* = 0,$$

$$\Sigma_i \theta_{i2} w_{i1}^* = -P, \ i = T, \ K, \ L, \tag{7}$$

where $P = p_1^* - p_2^*$, $w_{i1}^* = w_i^* - p_1^*$, $w_{i1} = w_i / p_1$; $P$ is the change in the relative price of a commodity; $w_{i1}$ is the real factor price measured by the price of good 1.

Totally differentiate eq. (3) to obtain

$$a_{ij}^* = \Sigma_h \varepsilon^{ij}_h w_h^*, \ i = T, \ K, \ L, \ j = 1, \ 2, \tag{8}$$

where

$$\varepsilon^{ij}_h = \partial \log a_{ij} / \partial \log w_h = \theta_{hj} \sigma^{ij}_h. \tag{9}$$

$\sigma^{ij}_h$ is the AES (or the Allen-partial elasticities of substitution) between the $i$th and the $h$th factors in the $j$th industry. For an additional definition of these symbols, see Sato and Koizumi (1973, p. 47-9), BC (p. 24). AESs are symmetric in the sense that

$$\sigma^{ij}_h = \sigma^{hj}_i. \tag{10}$$

According to BC (p. 33), 'Given the assumption that production functions are strictly quasi-concave and linearly homogeneous,'

$$\sigma^{ij}_i < 0. \tag{11}$$

Because $a_{ij}$ is homogeneous of degree zero in all input prices, we have

$$\Sigma_h \varepsilon^{ij}_h = \Sigma_h \theta_{hj} \sigma^{ij}_h = 0, \ i = T, \ K, \ L, \ j = 1, \ 2. \tag{12}$$

Eqs (8) to (12) are equivalent to the expressions in BC (p. 24, n.6). See also JE (p. 74, eqs (12)-(13)). From these, we derive

$$a_{ij}* = \Sigma_h \varepsilon^{ij}_h w_{h1}*. \tag{13}$$

Substituting eq. (13) in (4), we derive

$$\Sigma_j (\lambda_{ij} \Sigma_h \varepsilon^{ij}_h w_{h1}* + \lambda_{ij} X_j*) = \Sigma_h g_{ih} w_{h1}* + \Sigma_j \lambda_{ij} X_j* = V_i*, \ i = T, K, L. \tag{14}$$

where

$$g_{ih} = \Sigma_j \lambda_{ij} \varepsilon^{ij}_{h,} i, \ h = T, K, L. \tag{15}$$

This is the EWS (or the economy-wide substitution) between factors $i$ and $h$ defined by JE (p. 75). $g_{ih}$ is the aggregate of $\varepsilon^{ij}_h$. JE (p. 75) stated, 'Clearly, the substitution terms in the two industries are always averaged together. With this in mind, we define the term $\sigma^i_k$ to denote the economy-wide substitution towards or away from the use of factor $i$ when the $k$th factor becomes more expensive under the assumption that each industry's output is kept constant.'

We can easily show that

$$\Sigma_h g_{ih} = 0, i = T, K, L, \tag{16}$$

$$g_{ih} = (\theta_h / \theta_i) g_{hi}, i, \ h = T, K, L, \tag{17}$$

where $\theta_i$ and $\theta_j$ are, respectively, the share of factor $i, i = T, K, L$, and good $j, j = 1, 2$ in national income. That is, $\theta_j = p_j X_j / I$, $\theta_i = w_i V_i / I$, where $I = \Sigma_j p_j X_j = \Sigma_i w_i V_i$. See BC (p. 25, eq. (16)). Hence, we obtain $\lambda_{ij} = (\theta_j / \theta_i) \theta_{ij}$ (see JE (p. 72, n. 9)). Note that $\Sigma_j \theta_j = 1, \Sigma_i \theta_i = 1$. $g_{ih}$ is not symmetric. Specifically, $g_{ih} \neq g_{hi}, i \neq h$ in general. On eq. (17), see also JE (p. 85).

From (9), (11), and (15), we can show that

$$g_{ii} < 0. \tag{18}$$

From eqs (16) and (18), we derive

$$g_{KT} + g_{KL} = -g_{KK} > 0, \ g_{TK} + g_{TL} = -g_{TT} > 0, \ g_{LK} + g_{LT} = -g_{LL} > 0. \tag{19}$$

From (17) and (19), we can easily show that

$$(g_{LK}, g_{LT}, g_{KT}) = (+, +, +), (-, +, +), (+, -, +), (+, +, -). \tag{20}$$

At most, one of the EWSs $(g_{LK}, g_{LT}, g_{KT})$ can be negative.

Combine eqs (14) and (7) to make a system of linear equations. Using a 5 x 5 matrix, we obtain

$$\mathbf{AX} = \mathbf{P}, \qquad (21)$$

where $\mathbf{A} = \begin{bmatrix} \theta_{T1} & \theta_{K1} & \theta_{L1} & 0 & 0 \\ \theta_{T2} & \theta_{K2} & \theta_{L2} & 0 & 0 \\ g_{TT} & g_{TK} & g_{TL} & \lambda_{T1} & \lambda_{T2} \\ g_{KT} & g_{KK} & g_{KL} & \lambda_{K1} & \lambda_{K2} \\ g_{LT} & g_{LK} & g_{LL} & \lambda_{L1} & \lambda_{L2} \end{bmatrix}$, $\mathbf{X} = \begin{bmatrix} w_{T1}^* \\ w_{K1}^* \\ w_{L1}^* \\ X_1^* \\ X_2^* \end{bmatrix}$, $\mathbf{P} = \begin{bmatrix} 0 \\ -P \\ V_T^* \\ V_K^* \\ V_L^* \end{bmatrix}.$

$\mathbf{A}$ is a 5 x 5 coefficient matrix, and $\mathbf{X}, \mathbf{P}$ are column vectors.

2.2. Factor-intensity ranking

In this article, we assume

$$\frac{a_{T1}}{a_{T2}} > \frac{a_{L1}}{a_{L2}} > \frac{a_{K1}}{a_{K2}}. \qquad (22)$$

This implies

$$\frac{\theta_{T1}}{\theta_{T2}} > \frac{\theta_{L1}}{\theta_{L2}} > \frac{\theta_{K1}}{\theta_{K2}}. \qquad (23)$$

This is, 'the factor-intensity ranking' (see JE (p. 69), see also BC (p. 26-7), Suzuki (1983, p. 142)). This implies that sector 1 is relatively land-intensive, and sector 2 is relatively capital-intensive, labor is the middle factor, and land and capital are extreme factors (see also Ruffin (1981, p. 180)).

If (23) holds, we have

$$\frac{\theta_{L1}}{\theta_{L2}} > 1 \leftrightarrow \theta_{L1} > \theta_{L2}, \text{ or} \qquad (24)$$

$$\frac{\theta_{L1}}{\theta_{L2}} < 1 \leftrightarrow \theta_{L1} < \theta_{L2} \qquad (25)$$

Note that we do not assume that $\theta_{L1} = \theta_{L2}$ holds. JE (p. 70) called eqs (24) and (25) 'the factor-intensity ranking for the middle factor.' This implies that the middle factor is used relatively intensively in sector 1.

Define that

$$(A, B, E) = (\theta_{T1} - \theta_{T2}, \theta_{K1} - \theta_{K2}, \theta_{L1} - \theta_{L2}). \qquad (26)$$

This is the inter-sectoral difference in the distributional share. Recall (5) ($\Sigma_i \theta_{ij} = 1$), which implies

$$A + B + E = 0. \tag{27}$$

From eq. (27), we have

$$(A, B, E) = (-, +, -), (-, +, +), (+, +, -), (-, -, +), (+, -, +), (+, -, -). \tag{28}$$

However, eq. (23) implies

$$(A, B, E) = (+, -, ?). \tag{29}$$

From eqs (28) and (29), we have

$$(A, B, E) = (+, -, +), (+, -, -). \tag{30}$$

From eq. (27), for example, we obtain

$$E = -(A + B),$$

$$B = -(A + E). \tag{31}$$

If we assume eq. (24) holds, we derive

$$(A, B, E) = (+, -, +). \tag{32}$$

On the other hand, if we assume eq. (25) holds, we derive

$$(A, B, E) = (+, -, -). \tag{33}$$

In this article, we assume eqs (23) and (24) hold, hence, (32) holds.

### 2.3. EWS-ratio vector boundary

In this section, we derive the important relationship between EWS-ratios, and we draw the EWS-ratio vector boundary in the figure. This is useful for our analysis.

Each $a_{ij}$ function is homogeneous of degree zero in all input prices (see eq. (3)). According to BC (p. 33), 'Given the assumption that production functions are strictly quasi-concave and linearly homogeneous', $\sigma^{ij}_i < 0$ (see eq. (11)). These imply (see eq. (A17) in the Appendix A)

$$g_{KK} g_{TT} - g_{TK} g_{KT} > 0. \tag{34}$$

Recall eq. (19). That is, $g_{KK} = -(g_{KT} + g_{KL})$ and $g_{TT} = -(g_{TK} + g_{TL})$. Substitute these equations to eliminate $g_{KK}$ and $g_{TT}$ from L.H.S. of eq. (34). Next, recall eq. (17), that is, $g_{ih} = (\theta_h / \theta_i) g_{hi}$. Use this equation to eliminate $g_{KL}, g_{TL}$, and $g_{TK}$. That is, express using only three EWSs, namely, $g_{LK}$, $g_{LT}$, and $g_{KT}$:

$$\text{L.H.S. of (34)} = g_{KT}g_{TL} + g_{KL}g_{TK} + g_{KL}g_{TL} = \frac{\theta_L}{\theta_T}[g_{KT}(g_{LT} + g_{LK}) + \frac{\theta_L}{\theta_K} g_{LK}g_{LT}] \, (>0). \quad (35)$$

From eq. (19), $g_{LK} + g_{LT} = -g_{LL} > 0$. Using this, transform eq. (35) to obtain

$$g_{KT} > -\frac{\theta_L}{\theta_K} \frac{g_{LK}g_{LT}}{g_{LK} + g_{LT}}. \quad (36)$$

Define that, for ease of notation,

$$(S, T, U) = (g_{LK}, g_{LT}, g_{KT}). \quad (37)$$

If we use eq. (37), eq. (36) reduces

$$U > -\frac{\theta_L}{\theta_K} \frac{ST}{S+T}, \quad (38)$$

Dividing both sides of (38) by $T$, we derive

$$U' > -\frac{\theta_L}{\theta_K} \frac{S'}{S'+1}, \text{ if } T > 0; \; U' < -\frac{\theta_L}{\theta_K} \frac{S'}{S'+1}, \text{ if } T < 0, \quad (39)$$

where

$$(S', U') = (S/T, U/T) = (g_{LK}/g_{LT}, g_{KT}/g_{LT}), \quad (40)$$

which we call the EWS-ratio vector. $S'$ denotes the relative magnitude of EWS between factors $L$ and $K$ compared to EWS between factors $L$ and $T$. $U'$ denotes the relative magnitude of EWS between factors $K$ and $T$ compared to EWS between factors $L$ and $T$.

Transform

$$U' = -\frac{\theta_L}{\theta_K} \frac{S'}{S'+1} = -\frac{\theta_L}{\theta_K} + \frac{\theta_L}{\theta_K} \frac{1}{S'+1}, \quad (41)$$

which expresses the rectangular hyperbola. We call this the equation for the EWS-ratio vector boundary. It passes on the origin of $O$ (0, 0). The asymptotic lines are $S' = -1, \; U' = -\theta_L / \theta_K$. We can draw this boundary in the figure (see Fig. 1). $S'$ is written along the horizontal axis and $U'$ along the vertical axis. The EWS-ratio vector boundary demarcates the boundary of the region for the EWS-ratio vector. This implies that the EWS-ratio vector is not so arbitrary, but exists within these bounds.

Note that:

The EWS-ratio vector (S', U') exists in the upper-right region of the EWS-ratio vector boundary, if $T > 0$,

The EWS-ratio vector exists in the lower-left region of the EWS-ratio vector boundary if $T < 0$. (42)

The sign pattern of the EWS-ratio vector is, in each quadrant (on this, see also eq. (20)):

quad. I: $(S', U') = (+, +) \leftrightarrow (S, T, U) = (+, +, +)$;

quad. II: $(S', U') = (-, +) \leftrightarrow (S, T, U) = (-, +, +)$;

quad. III: $(S', U') = (-, -) \leftrightarrow (S, T, U) = (+, -, +)$;

quad. IV: $(S', U') = (+, -) \leftrightarrow (S, T, U) = (+, +, -)$. (43)

Hence, one of the EWS can be negative at most. Note that

$T > 0$, if (S', U') exists in quadrant I, II, or IV,

$T < 0$, if (S', U') exists in quadrant III, (44)

where we recall eqs (37) and (40), that is, $(S', U') = (S/T, U/T) = (g_{LK}/g_{LT}, g_{KT}/g_{LT})$, $(S, T, U) = (g_{LK}, g_{LT}, g_{KT})$. We define (for $i \neq h$),

Factors $i$ and $h$ are economy-wide substitutes, if $g_{ih} > 0$,

Factors $i$ and $h$ are economy-wide complements, if $g_{ih} < 0$. (45)

2.4. Solution

Using Cramer's rule to solve (21) for $X_2^*$, we derive

$$X_2^* = \Delta_5 / \Delta, \qquad (46)$$

where $\Delta = \det(\mathbf{A})$, $\Delta_5 = \det(\mathbf{A}_5) = \begin{vmatrix} \theta_{T1} & \theta_{K1} & \theta_{L1} & 0 & 0 \\ \theta_{T2} & \theta_{K2} & \theta_{L2} & 0 & -P \\ g_{TT} & g_{TK} & g_{TL} & \lambda_{T1} & V_T^* \\ g_{KT} & g_{KK} & g_{KL} & \lambda_{K1} & V_K^* \\ g_{LT} & g_{LK} & g_{LL} & \lambda_{L1} & V_L^* \end{vmatrix}$.

$\Delta$ is the determinant of matrix $\mathbf{A}$. We can show that $\Delta < 0$. On this, see Appendix B. Replacing column 5 of matrix $\mathbf{A}$ with column vector $\mathbf{P}$, we derive matrix $\mathbf{A}_5$. $\Delta_5$ is the determinant of matrix $\mathbf{A}_5$. Sum columns 1 and 2 in column 3, and subtract row 2 from row 1. We have

$$\Delta_5 = \begin{vmatrix} A & B & 0 & 0 & P \\ \theta_{T2} & \theta_{K2} & 1 & 0 & -P \\ g_{TT} & g_{TK} & 0 & \lambda_{T1} & V_T{}^* \\ g_{KT} & g_{KK} & 0 & \lambda_{K1} & V_K{}^* \\ g_{LT} & g_{LK} & 0 & \lambda_{L1} & V_L{}^* \end{vmatrix},$$

where we may recall eq. (26), that is, $(A, B, E) = (\theta_{T1} - \theta_{T2},\ \theta_{K1} - \theta_{K2}, \theta_{L1} - \theta_{L2})$. Express the above as a cofactor expansion along the third column:

$$\Delta_5 = (1)(-1)^{2+3} \begin{vmatrix} A & B & 0 & P \\ g_{TT} & g_{TK} & \lambda_{T1} & V_T{}^* \\ g_{KT} & g_{KK} & \lambda_{K1} & V_K{}^* \\ g_{LT} & g_{LK} & \lambda_{L1} & V_L{}^* \end{vmatrix}.$$

Express the above as a cofactor expansion along the fourth column:

$$\Delta_5 = (-1)^{2+3}\left[P(-1)^{1+4}C_{P2} + V_T{}^*(-1)^{2+4}C_{T2} + V_K{}^*(-1)^{3+4}C_{K2} + V_L{}^*(-1)^{4+4}C_{L2}\right] \quad (47)$$

where

$$C_{P2} = \begin{vmatrix} g_{TT} & g_{TK} & \lambda_{T1} \\ g_{KT} & g_{KK} & \lambda_{K1} \\ g_{LT} & g_{LK} & \lambda_{L1} \end{vmatrix},\ C_{T2} = \begin{vmatrix} A & B & 0 \\ g_{KT} & g_{KK} & \lambda_{K1} \\ g_{LT} & g_{LK} & \lambda_{L1} \end{vmatrix},\ C_{K2} = \begin{vmatrix} A & B & 0 \\ g_{TT} & g_{TK} & \lambda_{T1} \\ g_{LT} & g_{LK} & \lambda_{L1} \end{vmatrix},\ C_{L2} = \begin{vmatrix} A & B & 0 \\ g_{TT} & g_{TK} & \lambda_{T1} \\ g_{KT} & g_{KK} & \lambda_{K1} \end{vmatrix}.$$

(48)

From eqs (46) and (47), we have

$$X_2{}^* = \frac{\Delta_5}{\Delta} = \frac{1}{\Delta}(-1)\left[P(-C_{P2}) + V_T{}^* C_{T2} + V_K{}^*(-C_{K2}) + V_L{}^* C_{L2}\right]. \quad (49)$$

On the other hand, solve (21) for $X_1{}^*$:

$$X_1{}^* = \Delta_4/\Delta, \quad (50)$$

where $\Delta_4 = \det(\mathbf{A}_4) = \begin{vmatrix} \theta_{T1} & \theta_{K1} & \theta_{L1} & 0 & 0 \\ \theta_{T2} & \theta_{K2} & \theta_{L2} & -P & 0 \\ g_{TT} & g_{TK} & g_{TL} & V_T{}^* & \lambda_{T2} \\ g_{KT} & g_{KK} & g_{KL} & V_K{}^* & \lambda_{K2} \\ g_{LT} & g_{LK} & g_{LL} & V_L{}^* & \lambda_{L2} \end{vmatrix}.$

Replacing column 4 of matrix $\mathbf{A}$ with column vector $\mathbf{P}$, we derive the matrix $\mathbf{A}_4$. $\Delta_4$ is the determinant of matrix $\mathbf{A}_4$. Sum columns 1 and 2 in column 3, and subtract row 2 from row 1. We have

$$\Delta_4 = \begin{vmatrix} A & B & 0 & P & 0 \\ \theta_{T2} & \theta_{K2} & 1 & -P & 0 \\ g_{TT} & g_{TK} & 0 & V_T^* & \lambda_{T2} \\ g_{KT} & g_{KK} & 0 & V_K^* & \lambda_{K2} \\ g_{LT} & g_{LK} & 0 & V_L^* & \lambda_{L2} \end{vmatrix}.$$

Express the above as a cofactor expansion along the third column:

$$\Delta_4 = (1)(-1)^{2+3} \begin{vmatrix} A & B & P & 0 \\ g_{TT} & g_{TK} & V_T^* & \lambda_{T2} \\ g_{KT} & g_{KK} & V_K^* & \lambda_{K2} \\ g_{LT} & g_{LK} & V_L^* & \lambda_{L2} \end{vmatrix}.$$

Express the above as a cofactor expansion along the third column:

$$\Delta_4 = (-1)^{2+3}\left[ P(-1)^{1+3} C_{P1} + V_T^*(-1)^{2+3} C_{T1} + V_K^*(-1)^{3+3} C_{K1} + V_L^*(-1)^{4+3} C_{L1} \right] \quad (51)$$

where

$$C_{P1} = \begin{vmatrix} g_{TT} & g_{TK} & \lambda_{T2} \\ g_{KT} & g_{KK} & \lambda_{K2} \\ g_{LT} & g_{LK} & \lambda_{L2} \end{vmatrix},\ C_{T1} = \begin{vmatrix} A & B & 0 \\ g_{KT} & g_{KK} & \lambda_{K2} \\ g_{LT} & g_{LK} & \lambda_{L2} \end{vmatrix},\ C_{K1} = \begin{vmatrix} A & B & 0 \\ g_{TT} & g_{TK} & \lambda_{T2} \\ g_{LT} & g_{LK} & \lambda_{L2} \end{vmatrix},\ C_{L1} = \begin{vmatrix} A & B & 0 \\ g_{TT} & g_{TK} & \lambda_{T2} \\ g_{KT} & g_{KK} & \lambda_{K2} \end{vmatrix}, \quad (52)$$

Hence, from eqs (50) and (51), we have

$$X_1^* = \frac{\Delta_4}{\Delta} = \frac{1}{\Delta}(-1)[PC_{P1} + V_T^*(-C_{T1}) + V_K^* C_{K1} + V_L^*(-C_{L1})]. \quad (53)$$

In summary, from eqs (49) and (53), we obtain

$$X_2^* = \frac{1}{\Delta}[PC_{P2} + V_T^*(-C_{T2}) + V_K^* C_{K2} + V_L^*(-C_{L2})], \quad (54)$$

$$X_1^* = \frac{1}{\Delta}[P(-C_{P1}) + V_T^* C_{T1} + V_K^*(-C_{K1}) + V_L^* C_{L1}]. \quad (55)$$

2.5. Rybczynski matrix

From the above, the Rybczynski matrix $[X_j^*/V_i^*]$ (to use Thompson's terminology (1985, p. 619)) in elasticity terms is:

$$[X_j^*/V_i^*] = \begin{bmatrix} X_1^*/V_T^* & X_1^*/V_K^* & X_1^*/V_L^* \\ X_2^*/V_T^* & X_2^*/V_K^* & X_2^*/V_L^* \end{bmatrix} = \frac{1}{\Delta} \begin{bmatrix} C_{T1} & -C_{K1} & C_{L1} \\ -C_{T2} & C_{K2} & -C_{L2} \end{bmatrix}. \quad (56)$$

Express in general:

$$X_j^*/V_i^* = (1/\Delta)(-1)^{i+j} C_{ij}, \; i = T, K, L, \; j = 1, 2. \quad (57)$$

Substitute 1, 2, 3 instead of *T, K, L*, respectively, when we compute $(-1)^{i+j}$. Sign patterns are of interest. We can show that $X_1^*/V_L^*$ and $X_2^*/V_L^*$ are, respectively, equivalent to eq. (26) and (27) in BC (p. 32). BC only obtained these two equations. BC's method of derivation is somewhat complicated. The method shown here is simpler.

Using Saruss's rule to expand (48) and (52), we derive

$C_{T1} = A g_{KK} \lambda_{L2} + B \lambda_{K2} g_{LT} - (A g_{LK} \lambda_{K2} + B g_{KT} \lambda_{L2})$,

$C_{K1} = A g_{TK} \lambda_{L2} + B \lambda_{T2} g_{LT} - (A g_{LK} \lambda_{T2} + B g_{TT} \lambda_{L2})$,

$C_{L1} = A g_{TK} \lambda_{K2} + B \lambda_{T2} g_{KT} - (A g_{KK} \lambda_{T2} + B g_{TT} \lambda_{K2})$;

$C_{T2} = A g_{KK} \lambda_{L1} + B \lambda_{K1} g_{LT} - (A g_{LK} \lambda_{K1} + B g_{KT} \lambda_{L1})$,

$C_{K2} = A g_{TK} \lambda_{L1} + B \lambda_{T1} g_{LT} - (A g_{LK} \lambda_{T1} + B g_{TT} \lambda_{L1})$,

$C_{L2} = A g_{TK} \lambda_{K1} + B \lambda_{T1} g_{KT} - (A g_{KK} \lambda_{T1} + B g_{TT} \lambda_{K1})$. \quad (58)

Recall eq. (19), that is, $g_{KK} = -(g_{KT} + g_{KL})$ and $g_{TT} = -(g_{TK} + g_{TL})$. Substitute these equations into eq. (58) to eliminate $g_{KK}$ and $g_{TT}$. Next, recall eq. (17), that is, $g_{ih} = (\theta_h/\theta_i) g_{hi}$. Use this equation to eliminate $g_{KL}$, $g_{TL}$, and $g_{TK}$. Recall eq. (37), that is, $(S, T, U) = (g_{LK}, g_{LT}, g_{KT})$. Using these symbols for ease of notation, transform eq. (58):

$C_{T1} = E \lambda_{L2} U - A \dfrac{\theta_2}{\theta_K}(1-\theta_{T2})S + B \lambda_{K2} T,$

$C_{K1} = (-E) \dfrac{\theta_K}{\theta_T} \lambda_{L2} U - A \lambda_{T2} S + B \dfrac{\theta_2}{\theta_T}(1-\theta_{K2})T,$

$C_{L1} = (-E) \dfrac{\theta_2}{\theta_T}(1-\theta_{L2}) U + A \dfrac{\theta_L}{\theta_K} \lambda_{T2} S + B \dfrac{\theta_L}{\theta_T} \lambda_{K2} T;$

$C_{T2} = E \lambda_{L1} U - A \dfrac{\theta_1}{\theta_K}(1-\theta_{T1})S + B \lambda_{K1} T,$

$C_{K2} = (-E) \dfrac{\theta_K}{\theta_T} \lambda_{L1} U - A \lambda_{T1} S + B \dfrac{\theta_1}{\theta_T}(1-\theta_{K1})T,$

$$C_{L2} = (-E)\frac{\theta_1}{\theta_T}(1-\theta_{L1})U + A\frac{\theta_L}{\theta_K}\lambda_{T1}S + B\frac{\theta_L}{\theta_T}\lambda_{K1}T, \qquad (59)$$

where we recall eq. (31), that is, $E = -(A+B)$. $C_{ij}$ is a linear function in $S$, $T$, and $U$.   Recall eq. (40), that is, $(S',U') = (S/T, U/T) = (g_{LK}/g_{LT}, g_{KT}/g_{LT})$, which we call the EWS-ratio vector. Using this, transform eq. (59) to derive

$$C_{T1} = E\lambda_{L2} T[U'-f_{T1}(S')]], \quad C_{K1} = (-E)\frac{\theta_K}{\theta_T}\lambda_{L2} T[U'-f_{K1}(S')],$$

$$C_{L1} = (-E)\frac{\theta_2}{\theta_T}(1-\theta_{L2}) T[U'-f_{L1}(S')];$$

$$C_{T2} = E\lambda_{L1} T[U'-f_{T2}(S')]], \quad C_{K2} = (-E)\frac{\theta_K}{\theta_T}\lambda_{L1} T[U'-f_{K2}(S')],$$

$$C_{L2} = (-E)\frac{\theta_1}{\theta_T}(1-\theta_{L1})T[U'-f_{L2}(S')], \qquad (60)$$

where

$$f_{T1}(S') = [A\frac{\theta_2}{\theta_K}(1-\theta_{T2})S' - B\lambda_{K2}](E\lambda_{L2})^{-1},$$

$$f_{K1}(S') = [A\lambda_{T2}S' - B\frac{\theta_2}{\theta_T}(1-\theta_{K2})][(-E)\frac{\theta_K}{\theta_T}\lambda_{L2}]^{-1},$$

$$f_{L1}(S') = [-A\frac{\theta_L}{\theta_K}\lambda_{T2}S' - B\frac{\theta_L}{\theta_T}\lambda_{K2}][(-E)\frac{\theta_2}{\theta_T}(1-\theta_{L2})]^{-1};$$

$$f_{T2}(S') = [A\frac{\theta_1}{\theta_K}(1-\theta_{T1})S' - B\lambda_{K1}](E\lambda_{L1})^{-1},$$

$$f_{K2}(S') = [A\lambda_{T1}S' - B\frac{\theta_1}{\theta_T}(1-\theta_{K1})][(-E)\frac{\theta_K}{\theta_T}\lambda_{L1}]^{-1},$$

$$f_{L2}(S') = [-A\frac{\theta_L}{\theta_K}\lambda_{T1}S' - B\frac{\theta_L}{\theta_T}\lambda_{K1}][(-E)\frac{\theta_1}{\theta_T}(1-\theta_{L1})]^{-1}. \qquad (61)$$

Define

$$f_{ij}(S') = [A_{ij}S' + B_{ij}]E_{ij}^{-1}, \text{ and } C_{ij}' = U' - f_{ij}(S'), i = T, K, L, j = 1, 2. \qquad (62)$$

In these expressions, $A_{ij}$, $B_{ij}$, and $E_{ij}$ are the parameters respectively related to $A$, $B$, and $E$. That is,

$$(A_{ij}, \quad B_{ij}, \quad E_{ij}) = (A\frac{\theta_2}{\theta_K}(1-\theta_{T2}), -B\lambda_{K2}, E\lambda_{L2}), \text{ for } ij = T1,$$

$$= (A\lambda_{T2}, -B\frac{\theta_2}{\theta_T}(1-\theta_{K2}), (-E)\frac{\theta_K}{\theta_T}\lambda_{L2}), \text{ for } ij = K1,$$

$$= (-A\frac{\theta_L}{\theta_K}\lambda_{T2}, -B\frac{\theta_L}{\theta_T}\lambda_{K2}, (-E)\frac{\theta_2}{\theta_T}(1-\theta_{L2})), \text{ for } ij = L1;$$

$$= (A\frac{\theta_1}{\theta_K}(1-\theta_{T1}), -B\lambda_{K1}, E\lambda_{L1}), \text{ for } ij = T2,$$

$$= (A\lambda_{T1}, -B\frac{\theta_1}{\theta_T}(1-\theta_{K1}), (-E)\frac{\theta_K}{\theta_T}\lambda_{L1}), \text{ for } ij = K2,$$

$$= (-A\frac{\theta_L}{\theta_K}\lambda_{T1}, -B\frac{\theta_L}{\theta_T}\lambda_{K1}, (-E)\frac{\theta_1}{\theta_T}(1-\theta_{L1})), \text{ for } ij = L2. \tag{63}$$

$C_{ij}'$ is a linear function in $S'$ and $U'$. Express in general:

$$C_{ij} = E_{ij}T\ C_{ij}', i = T, K, L, j = 1, 2. \tag{64}$$

2.6. Drawing the border line for a Rybczynski sign pattern to change

From eqs (57), (64), and (62), we derive

$$X_j^*/V_i^* = \frac{1}{\Delta}(-1)^{i+j}C_{ij} = \frac{1}{\Delta}(-1)^{i+j}E_{ij}T\ C_{ij}' = \frac{1}{\Delta}(-1)^{i+j}E_{ij}T[U' - f_{ij}(S')]. \tag{65}$$

From eq. (65), we derive

$$X_j^*/V_i^* = 0 \leftrightarrow C_{ij} = 0 \leftrightarrow C_{ij}' = 0$$

$$\leftrightarrow U' = f_{ij}(S') = [A_{ij}S' + B_{ij}]E_{ij}^{-1}, i = T, K, L, j = 1, 2. \tag{66}$$

This equation expresses the straight line in two dimensions. We call it the equation for line $ij$, which expresses the border line for a Rybczynski sign pattern to change. The gradient and intercept of line $ij$ are, respectively, $A_{ij}E_{ij}^{-1}$ and $B_{ij}E_{ij}^{-1}$.

Using eqs (66) and (41), make a system of equations:

$$U' = f_{ij}(S') = [A_{ij}S' + B_{ij}]E_{ij}^{-1}, i = T, K, L, j = 1, 2, \tag{67}$$

$$U' = -\frac{\theta_L}{\theta_K}\frac{S'}{S'+1}. \tag{68}$$

From these, we obtain a quadratic equation in $S'$ for each $i, j$. Solve this to derive two solutions. Each solution denotes the $S'$ coordinate value of the intersection point of line $ij$ and EWS-ratio vector boundary. The solutions are, for line-*T1, K1, L1; T2, K2, L2*, respectively:

$$S' = \frac{B}{A}, \frac{-\theta_{K2}}{1-\theta_{T2}}, \quad S' = \frac{B}{A}, \frac{-(1-\theta_{K2})}{\theta_{T2}}, \quad S' = \frac{B}{A}, \frac{\theta_{K2}}{\theta_{T2}};$$

$$S' = \frac{B}{A}, \frac{-\theta_{K1}}{1-\theta_{T1}}, \quad S' = \frac{B}{A}, \frac{-(1-\theta_{K1})}{\theta_{T1}}, \quad S' = \frac{B}{A}, \frac{\theta_{K1}}{\theta_{T1}}. \tag{69}$$

In summary, there are seven intersection points. Each line $ij$ passes through the same point, which we call point $Q$. The Cartesian coordinates of point $Q$ are

$$(S', U') = (\frac{B}{A}, \frac{B}{E}\frac{\theta_L}{\theta_K}). \tag{70}$$

We call six intersection points other than point $Q$, the point $R_{ij}, i=T, K, L, j=1, 2$. The Cartesian coordinates of these points are, for line-*T1, K1, L1; T2, K2, L2*, respectively:

$$(S', U') = (\frac{-\theta_{K2}}{1-\theta_{T2}}, \frac{\theta_{K2}}{\theta_{L2}}\frac{\theta_L}{\theta_K}), (\frac{-(1-\theta_{K2})}{\theta_{T2}}, \frac{1-\theta_{K2}}{-\theta_{L2}}\frac{\theta_L}{\theta_K}), (\frac{\theta_{K2}}{\theta_{T2}}, \frac{-\theta_{K2}}{1-\theta_{L2}}\frac{\theta_L}{\theta_K});$$

$$(\frac{-\theta_{K1}}{1-\theta_{T1}}, \frac{\theta_{K1}}{\theta_{L1}}\frac{\theta_L}{\theta_K}), (\frac{-(1-\theta_{K1})}{\theta_{T1}}, \frac{1-\theta_{K1}}{-\theta_{L1}}\frac{\theta_L}{\theta_K}), (\frac{\theta_{K1}}{\theta_{T1}}, \frac{-\theta_{K1}}{1-\theta_{L1}}\frac{\theta_L}{\theta_K}). \tag{71}$$

From eqs (23) and (24), we derive eq. (32), that is, $(A, B, E) = (+, -, +)$. Substituting this in eq. (70), we derive the sign pattern of point $Q$, that is,

$$\text{sign}(S', U') = (-, -). \tag{72}$$

This implies that point $Q$ belongs to quadrant III.

The sign patterns of point $R_{ij}$ are, respectively,

$$\text{sign}(S', U') = (-, +), (-, -), (+, -); (-, +), (-, -), (+, -). \tag{73}$$

Hence, point $R_{T1}$ and $R_{T2}$ are in quadrant II; point $R_{K1}$ and $R_{K2}$ are in quadrant III; and point $R_{L1}$ and $R_{L2}$ are in quadrant IV.

Next, we investigate the relative position of point $R_{ij}$ and $Q$. From eq. (23), we can prove for $S'$ values of point $R_{K1}$ and $R_{K2}$:

$$\frac{-(1-\theta_{K2})}{\theta_{T2}} < \frac{-(1-\theta_{K1})}{\theta_{T1}}. \tag{74}$$

Equation (74) explains the relative position of the two points ($R_{K1}$ and $R_{K2}$). Similarly, from eq. (23), we can prove for $S'$ values of point $R_{T1}$, $R_{T2}$, the origin of $O$, point $R_{L2}$, and $R_{L1}$:

$$\frac{-\theta_{K2}}{1-\theta_{T2}} < \frac{-\theta_{K1}}{1-\theta_{T1}} < 0 < \frac{\theta_{K1}}{\theta_{T1}} < \frac{\theta_{K2}}{\theta_{T2}}. \tag{75}$$

Equation (75) explains the relative position of these five points.

We can prove for $S'$ values of point $R_{K2}$ and $Q$:

$$\frac{-(1-\theta_{K1})}{\theta_{T1}} < \frac{B}{A}. \tag{76}$$

The derivation of eq. (76) is as follows. Because we assume (32), that is, $(A, B, E) = (+, -, +)$, we have $A = (+)$. Hence, (76) reduces

$$-(1-\theta_{K1})A < \theta_{T1}B \tag{77}$$

Recall eq. (31), $B = -(A+E)$. Substitute this in eq. (77) and multiply both sides by (-1). By transforming this, we derive

$$\theta_{L1}A - E\theta_{T1} > 0 \tag{78}$$

Using eq. (23), we can show that

$$L.H.S. \text{ of } (78) = \theta_{L1}(\theta_{T1}-\theta_{T2}) - (\theta_{L1}-\theta_{L2})\theta_{T1} = -\theta_{L1}\theta_{T2} + \theta_{L2}\theta_{T1} > 0.$$

Thus, we have proved eq. (76).

From eqs (70)-(76), we can draw point $Q$ and $R_{ij}$ and, hence, line $ij$ in the figure. Each line $ij$ divides the region of the EWS-ratio vector into 12 subregions, that is, the subregion $P$1-5 (upper-right region) and $M$1-7 (lower-left region) (see Fig. 1).

2.7. Rybczynski sign patterns

Define the 2 x 3 matrices:

$$\mathbf{F} = \left[F_{ij}\right] = \begin{bmatrix} 1 & -1 & 1 \\ -1 & 1 & -1 \end{bmatrix}, \quad \mathbf{C} = \left[C_{ij}\right] = \begin{bmatrix} C_{T1} & C_{K1} & C_{L1} \\ C_{T2} & C_{K2} & C_{L2} \end{bmatrix},$$

$$\mathbf{E} = [E_{ij}] = \begin{bmatrix} E_{T1} & E_{K1} & E_{L1} \\ E_{T2} & E_{K2} & E_{L2} \end{bmatrix} = \begin{bmatrix} E\lambda_{L2} & (-E)\dfrac{\theta_K}{\theta_T}\lambda_{L2} & (-E)\dfrac{\theta_2}{\theta_T}(1-\theta_{L2}) \\ E\lambda_{L1} & (-E)\dfrac{\theta_K}{\theta_T}\lambda_{L1} & (-E)\dfrac{\theta_1}{\theta_T}(1-\theta_{L1}) \end{bmatrix},$$

$$\mathbf{C'} = [C_{ij}'] = \begin{bmatrix} C_{T1}' & C_{K1}' & C_{L1}' \\ C_{T2}' & C_{K2}' & C_{L2}' \end{bmatrix} = \begin{bmatrix} U' - f_{T1}(S') & U' - f_{K1}(S') & U' - f_{L1}(S') \\ U' - f_{T2}(S') & U' - f_{K2}(S') & U' - f_{L2}(S') \end{bmatrix}. \quad (79)$$

Using the Hadamard product of these matrices, we can transform eq. (56):

$$[X_j*/V_i*] = \frac{1}{\Delta}\mathbf{F} \circ \mathbf{C}, \quad (80)$$

where (see eq. (64))

$$\mathbf{C} = \mathbf{E} \circ \mathbf{C'}T. \quad (81)$$

In general, if $\mathbf{A} = [a_{ij}]$ and $\mathbf{B} = [b_{ij}]$ are each $m \times n$ matrices, their Hadamard product is the matrix of element-wise products, that is, $\mathbf{A} \circ \mathbf{B} = [a_{ij}b_{ij}]$. For this definition, see, for example, Styan (1973, p. 217-18). Hadamard product is known, for example, in the literature of statistics.

Hence, Rybczynski sign patterns are:

$$sign[X_j*/V_i*] = sign\frac{1}{\Delta}\mathbf{F} \circ \mathbf{C} = sign\frac{1}{\Delta}\mathbf{F} \circ sign\mathbf{C}, \quad (82)$$

where

$$sign\mathbf{C} = sign(\mathbf{E} \circ \mathbf{C'}T) = sign\mathbf{E} \circ sign\mathbf{C'}T. \quad (83)$$

Recall that $\Delta < 0$ (see eq. (46)). Hence,

$$sign\frac{1}{\Delta}\mathbf{F} = sign\frac{1}{\Delta}sign\begin{bmatrix} 1 & -1 & 1 \\ -1 & 1 & -1 \end{bmatrix} = (-)\begin{bmatrix} + & - & + \\ - & + & - \end{bmatrix} = \begin{bmatrix} - & + & - \\ + & - & + \end{bmatrix}. \quad (84)$$

Recall that we assume eq. (32), that is, $(A, B, E) = (+, -, +)$. Hence,

$$sign\mathbf{E} = \begin{bmatrix} + & - & - \\ + & - & - \end{bmatrix}. \quad (85)$$

In general, if the EWS-ratio vector ($S'$, $U'$) exists in the subregion above line $ij$ (resp. below line $ij$), we derive

$$C_{ij}' = U' - f_{ij}(S') = (+) > 0. \quad (resp. C_{ij}' = U' - f_{ij}(S') = (-) < 0). \quad (86)$$

For example, if the EWS-ratio vector exists in subregion P2, that is, below line T1, T2, L2, and above line L1, K1, K2, the sign pattern of matrix $\mathbf{C'}$ is

$$sign\mathbf{C}" = \begin{bmatrix} C_{ij}' \end{bmatrix} = \begin{bmatrix} - & + & + \\ - & + & - \end{bmatrix}.$$

Sign patterns of matrix **C'** are, respectively, for each subregion:

$$sign\mathbf{C'} = \overset{P1}{\begin{bmatrix} - & + & - \\ - & + & - \end{bmatrix}} \overset{P2}{\begin{bmatrix} - & + & + \\ - & + & - \end{bmatrix}} \overset{P3}{\begin{bmatrix} - & + & + \\ - & + & + \end{bmatrix}} \overset{P4}{\begin{bmatrix} - & + & + \\ + & + & + \end{bmatrix}} \overset{P5}{\begin{bmatrix} + & + & + \\ + & + & + \end{bmatrix}}$$

$$\overset{M1}{\begin{bmatrix} + & + & + \\ + & + & + \end{bmatrix}} \overset{M2}{\begin{bmatrix} + & + & + \\ + & - & + \end{bmatrix}} \overset{M3}{\begin{bmatrix} + & - & + \\ + & - & + \end{bmatrix}} \overset{M4}{\begin{bmatrix} + & - & - \\ + & - & + \end{bmatrix}} \overset{M5}{\begin{bmatrix} + & - & - \\ + & - & - \end{bmatrix}} \overset{M6}{\begin{bmatrix} + & - & - \\ - & - & - \end{bmatrix}} \overset{M7}{\begin{bmatrix} - & - & - \\ - & - & - \end{bmatrix}} \quad (87)$$

In summary, the position of the EWS-ratio vector determines the sign pattern of matrix **C'**.

Of course, we can state that from eq. (44) and Fig. 1,

$T > 0$, if the EWS-ratio vector exists in any of the subregions P1-P5,

$T < 0$, if the EWS-ratio vector exists in any of the subregions M1-M7. (88)

From eqs (87) and (88), the sign patterns of the matrix **C'**$T$ are, for each subregion:

$$sign\mathbf{C'} = \overset{P1}{\begin{bmatrix} - & + & - \\ - & + & - \end{bmatrix}} \overset{P2}{\begin{bmatrix} - & + & + \\ - & + & - \end{bmatrix}} \overset{P3}{\begin{bmatrix} - & + & + \\ - & + & + \end{bmatrix}} \overset{P4}{\begin{bmatrix} - & + & + \\ + & + & + \end{bmatrix}} \overset{P5}{\begin{bmatrix} + & + & + \\ + & + & + \end{bmatrix}}$$

$$\overset{M1}{\begin{bmatrix} - & - & - \\ - & - & - \end{bmatrix}} \overset{M2}{\begin{bmatrix} - & - & - \\ - & + & - \end{bmatrix}} \overset{M3}{\begin{bmatrix} - & + & - \\ - & + & - \end{bmatrix}} \overset{M4}{\begin{bmatrix} - & + & + \\ - & + & - \end{bmatrix}} \overset{M5}{\begin{bmatrix} - & + & + \\ - & + & + \end{bmatrix}} \overset{M6}{\begin{bmatrix} - & + & + \\ + & + & + \end{bmatrix}} \overset{M7}{\begin{bmatrix} + & + & + \\ + & + & + \end{bmatrix}} \quad (89)$$

Note that the sign patterns for P1-P5 are, respectively, the same as those for M3-M7.

Recall eq. (83), that is, $sign\mathbf{C} = sign(\mathbf{E} \circ \mathbf{C'}T) = sign\mathbf{E} \circ sign\mathbf{C'}T$. Substituting eqs (89) and (85) in (83), we have

$$sign\mathbf{C} = \overset{P1}{\begin{bmatrix} - & - & + \\ - & - & + \end{bmatrix}} \overset{P2}{\begin{bmatrix} - & - & - \\ - & - & + \end{bmatrix}} \overset{P3}{\begin{bmatrix} - & - & - \\ - & - & - \end{bmatrix}} \overset{P4}{\begin{bmatrix} - & - & - \\ + & - & - \end{bmatrix}} \overset{P5}{\begin{bmatrix} + & - & - \\ + & - & - \end{bmatrix}}$$

$$\overset{M1}{\begin{bmatrix} - & + & + \\ - & + & + \end{bmatrix}} \overset{M2}{\begin{bmatrix} - & + & + \\ - & - & + \end{bmatrix}} \overset{M3}{\begin{bmatrix} - & - & + \\ - & - & + \end{bmatrix}} \overset{M4}{\begin{bmatrix} - & - & - \\ - & - & + \end{bmatrix}} \overset{M5}{\begin{bmatrix} - & - & - \\ - & - & - \end{bmatrix}} \overset{M6}{\begin{bmatrix} - & - & - \\ + & - & - \end{bmatrix}} \overset{M7}{\begin{bmatrix} + & - & - \\ + & - & - \end{bmatrix}} \quad (90)$$

Recall eq. (82), that is, $sign[X_j^*/V_i^*] = sign\frac{1}{\Delta}\mathbf{F} \circ sign\mathbf{C}$. Substitute eqs (90) and (84) in (82), we derive the Rybczynski sign patterns. They are, for each subregion:

$$\text{sign}\left[X_j^*/V_i^*\right] = \overset{P1}{\begin{bmatrix} + & - & - \\ - & + & + \end{bmatrix}} \overset{P2}{\begin{bmatrix} + & - & + \\ - & + & + \end{bmatrix}} \overset{P3}{\begin{bmatrix} + & - & + \\ - & + & - \end{bmatrix}} \overset{P4}{\begin{bmatrix} + & - & + \\ + & + & - \end{bmatrix}} \overset{P5}{\begin{bmatrix} - & - & + \\ + & + & - \end{bmatrix}}$$

$$\overset{M1}{\begin{bmatrix} + & + & - \\ - & - & + \end{bmatrix}} \overset{M2}{\begin{bmatrix} + & + & - \\ - & + & + \end{bmatrix}} \overset{M3}{\begin{bmatrix} + & - & - \\ - & + & + \end{bmatrix}} \overset{M4}{\begin{bmatrix} + & - & + \\ - & + & + \end{bmatrix}} \overset{M5}{\begin{bmatrix} + & - & + \\ - & + & - \end{bmatrix}} \overset{M6}{\begin{bmatrix} + & - & + \\ + & + & - \end{bmatrix}} \overset{M7}{\begin{bmatrix} - & - & + \\ + & + & - \end{bmatrix}} \quad (91)$$

In summary, the position of the EWS-ratio vector determines the Rybczynski sign pattern. There are 12 patterns in total. Note that the sign patterns for P1-P5 are, respectively, the same as those for M3-M7. If we do not count the duplication, there are seven patterns in total.

We make the following statements.

(i) If the EWS ratio vector (*S'*, U') exists in subregion P1, P2, or P3, the effects of land endowment on commodity output in sector 1 and sector 2 are positive and negative, respectively. The effects of capital endowment on commodity output in sector 1 and sector 2 are negative and positive, respectively.

(ii) If the EWS ratio vector exists in subregion P4 or M6, the effects of land endowment on commodity output in both sectors 1 and 2 are positive. The effects of capital endowment on commodity output in sector 1 and sector 2 are negative and positive, respectively.

(iii) If the EWS ratio vector exists in subregion P5 or M7, the effects of land endowment on commodity output in sector 1 and sector 2 are negative and positive, respectively. The effects of capital endowment on commodity output in sector 1 and sector 2 are negative and positive, respectively.

(iv) If the EWS ratio vector exists in subregion M1, the effects of land endowment on commodity output in sector 1 and sector 2 are positive and negative, respectively. The effects of capital endowment on commodity output in sector 1 and sector 2 are positive and negative, respectively.

(v) If the EWS ratio vector exists in Subregion M2, the effects of land endowment on commodity output in sector 1 and sector 2 are positive and negative, respectively. The effects of capital endowment on commodity output in both sectors 1 and 2 are positive.

We can state as follows.

A strong Rybczynski result holds if the EWS-ratio vector exists in the subregion P1, P2, P3; M3, M4, or M5.

A strong Rybczynski result does not hold if the EWS-ratio vector exists in the subregion P4, P5; M1, M2, M6, or M7.

(92)

2.8. The commodity price–factor price relationship

From the reciprocity relations derived by Samuelson, BC (p. 36, eqs (31)-(33)) derived

$$\frac{w_i{}^* - p_1{}^*}{P} = \frac{X_2{}^*}{V_i{}^*} \frac{-\theta_2}{\theta_i},$$

$$\frac{w_i{}^* - p_2{}^*}{P} = \frac{X_1{}^*}{V_i{}^*} \frac{\theta_1}{\theta_i}, i = T, K, L, \quad (93)$$

where we recall (7), that is, $P = p_1{}^* - p_2{}^*$. Define the Stolper-Samuelson matrix in elasticity terms:

$$\left[\frac{w_i{}^* - p_j{}^*}{P}\right] = \begin{bmatrix} w_T{}^* - p_1{}^* & w_K{}^* - p_1{}^* & w_L{}^* - p_1{}^* \\ w_T{}^* - p_2{}^* & w_K{}^* - p_2{}^* & w_L{}^* - p_2{}^* \end{bmatrix} \frac{1}{P}. \quad (94)$$

This matrix shows how the relative price of a commodity affects the real factor prices. Sign patterns are of interest. Multiply the second row of eq. (91) by (-1) and interchange row 1 and row 2, we derive the Stolper-Samuelson sign patterns as follows. They are, for each subregion:

$$\text{sign}\left[\frac{w_i{}^* - p_j{}^*}{P}\right] = \overset{P1}{\begin{bmatrix} + & - & - \\ + & - & - \end{bmatrix}} \overset{P2}{\begin{bmatrix} + & - & - \\ + & - & + \end{bmatrix}} \overset{P3}{\begin{bmatrix} + & - & + \\ + & - & + \end{bmatrix}} \overset{P4}{\begin{bmatrix} - & - & + \\ + & - & + \end{bmatrix}} \overset{P5}{\begin{bmatrix} - & - & + \\ - & - & + \end{bmatrix}}$$

$$\overset{M1}{\begin{bmatrix} + & + & - \\ + & + & - \end{bmatrix}} \overset{M2}{\begin{bmatrix} + & - & - \\ + & + & - \end{bmatrix}} \overset{M3}{\begin{bmatrix} + & - & - \\ + & - & - \end{bmatrix}} \overset{M4}{\begin{bmatrix} + & - & - \\ + & - & + \end{bmatrix}} \overset{M5}{\begin{bmatrix} + & - & + \\ + & - & + \end{bmatrix}} \overset{M6}{\begin{bmatrix} - & - & + \\ + & - & + \end{bmatrix}} \overset{M7}{\begin{bmatrix} - & - & + \\ - & - & + \end{bmatrix}} \quad (95)$$

In summary, the position of the EWS-ratio vector determines the Stolper-Samuelson sign pattern.

Note that

the sign patterns of matrix $[w_i{}^* - p_j{}^*]$ are similar to eq. (95), if $P = (+) > 0$,

the sign patterns of matrix $[w_i{}^* - p_j{}^*]$ are opposite to eq. (95), if $P = (-) < 0$. (96)

3. Some applications

Example 1: For example, we derive the following.

(i) If $(S', U') = (+, +)$, the EWS-ratio vector exists in quadrant I, that is, in the subregion P1-P5.

(ii) If $(S', U') = (-, +)$, the EWS-ratio vector exists in quadrant II, that is, in the subregion P3, P4, or P5.

(iii) If $(S', U') = (-, -)$, the EWS-ratio vector exists in quadrant III, that is, in the subregion M1-M7.

In all three cases, it is indeterminate whether a strong Rybczynski result holds.

(iv) If $(S', U') = (+, -)$, the EWS-ratio vector exists in quadrant IV, that is, in the subregion P1, P2, or P3.

We assume (iv) holds. That is (see eq. (43)),

$$(S', U') = (S/T, U/T) = (+, -)$$
$$\leftrightarrow (S, T, U) = (g_{LK}, g_{LT}, g_{KT}) = (+, +, -). \quad (97)$$

This implies that capital and land, extreme factors, are economy-wide complements. From (92), a strong Rybczynski result holds necessarily. Hence, the Rybczynski sign patterns for P1-P3 hold (see (91)). The following result has been established.

*Theorem 1*. We assume the factor-intensity ranking as follows.

$$\frac{\theta_{T1}}{\theta_{T2}} > \frac{\theta_{L1}}{\theta_{L2}} > \frac{\theta_{K1}}{\theta_{K2}}, \quad (98)$$

$$\frac{\theta_{L1}}{\theta_{L2}} > 1 \leftrightarrow \theta_{L1} > \theta_{L2}. \quad (99)$$

Further, if the EWS-ratio vector $(S', U')$ exists in quadrant IV (or subregions P1-P3), in other words, if capital and land, extreme factors, are economy-wide complements, a strong Rybczynski result necessarily holds. In this case, the Rybczynski sign patterns, as per Thompson's (1985, p. 619) terminology for subregions P1-P3 are, respectively:

$$\text{sign}[X_j^*/V_i^*] = \begin{matrix} P1 & P2 & P3 \end{matrix}$$
$$\begin{bmatrix} + & - & - \\ - & + & + \end{bmatrix} \begin{bmatrix} + & - & + \\ - & + & + \end{bmatrix} \begin{bmatrix} + & - & + \\ - & + & - \end{bmatrix}. \quad (100)$$

The Stolper-Samuelson sign patterns for subregions P1-P3 are, respectively:

$$\text{sign}[\frac{w_i^* - p_j^*}{P}] = \begin{bmatrix} + & - & - \\ + & - & - \end{bmatrix} \begin{bmatrix} + & - & - \\ + & - & + \end{bmatrix} \begin{bmatrix} + & - & + \\ + & - & + \end{bmatrix}. \quad (101)$$

Eq. (100) implies that each sign pattern expresses the factor endowment–commodity output relationship. Notably, the sign of Column 3 shows the labor endowment–commodity output relationship. An increase in the supply of land increases the output of commodity 1 and reduces the output of commodity 2. Moreover, an increase in the supply of capital increases the output of commodity 2 and reduces the output of commodity 1. However, it is indeterminate how an increase in the supply of labor affects the outputs of commodities 1 and 2. Three patterns are possible. Therefore, we make the following statements.

(i) If the EWS ratio vector ($S'$, $U'$) exists in subregion P1, the effects of labor endowment on commodity output in sector 1 and sector 2 are negative and positive, respectively.

(ii) If the EWS ratio vector exists in subregion P2, the effects of labor endowment on commodity output in both sectors 1 and 2 are positive.

(iii) If the EWS ratio vector exists in subregion P3, the effects of labor endowment on commodity output in sector 1 and sector 2 are positive and negative, respectively.

Eq. (101) implies as follows. Each sign pattern expresses the commodity price–factor price relationships. For example, if we assume that P = (+) > 0, the sign patterns of the matrix [ $w_i^* - p_j^*$ ] are similar to the above. That is, both the real factor prices of land measured by good 1 and 2 increase, and both the real factor prices of capital decrease.

(i) If the EWS-ratio vector ($S'$, $U'$) exists in the subregion P1, both the real factor prices of labor measured by good 1 and 2 decrease. This is not favorable for the labor owner.

(ii) If the EWS-ratio vector exists in subregion P2, the real factor price of labor measured by good 1 decreases, and the real factor price of labor measured by good 2 increases. It is indeterminate whether this is favorable for the labor owner.

(iii) If the EWS-ratio vector exists in subregion P3, both the real factor prices of labor measured by good 1 and 2 increase. This is favorable for the labor owner.

On the other hand, if we assume P = (-) < 0, the sign patterns of the matrix [ $w_i^* - p_j^*$ ] are the opposite to the above. For example, as in Takayama (1982, p. 20), we can apply these results to the US trade problem in the 1980s. Takayama (1982) did not analyze in elasticity terms but in differential forms. If we replace factors T, K, L in our analysis with factors 1, 2, 3, respectively, Takayama's (1982) result is very similar to ours.

Takayama (1982) calls factors 1, 2, and 3, respectively, skilled labor, (physical) capital, and unskilled labor (or raw labor). The author called industries 1 and 2, respectively, exportable and importable.

The author also states, 'there seems to be strong evidence that the current US commodity structure of trade is such that her exports are relatively skilled labor (or R&D) intensive vis-a-vis unskilled labor, and that her imports are relatively capital intensive vis-a-vis unskilled labor (e.g., Baldwin, 1971, 1979).' This implies (see Takayama (1982, p. 14, p. 20))

$$\frac{a_{11}}{a_{12}} > \frac{a_{31}}{a_{32}} > \frac{a_{21}}{a_{22}}. \qquad (102)$$

This is the factor-intensity ranking. Takayama (1982, p. 20) continues, 'there is some evidence that skilled labor and capital are (aggregate) complements (e.g., Branson-Monoyios, 1977). This indicates that our assumption of [aggregate] complements for extreme factors are satisfied.'

This implies that $s_{12} < 0$ (see Takayama (p. 18)). This implies $g_{12} < 0$, if we use EWS. The reason is that $s_{ih} = g_{ih} V_i / w_h$, $i, h = 1, 2, 3$ (see (A16)). Hence, the EWS-ratio vector exists in quadrant IV, that is, in either of the subregions P1, P2, or P3.

Takayama (1982) derived the sign pattern of 'the Stolper-Samuelson matrix' (see Takayama (p. 20)). If we use our symbols, the sign pattern is:

$$\left[\partial X_j / \partial V_i\right] = \left[\partial w_i / \partial p_j^*\right]^t = \begin{bmatrix} + & - & ? \\ - & + & ? \end{bmatrix}, \qquad (103)$$

where t denotes the transpose. Takayama (1982, p. 20) states, 'we may conclude that an import restriction which raises the domestic price of importables (say, automobiles from Japan) in the US increases the return on capital and lowers the return on skilled labor (or R&D) in the US.' Similarly, the author analyzed the effect of a reduction on import restrictions, which is the opposite of the above.

Takayama (1982) only analyzed the effect on the price of extreme factors (factors 1 and 2). He did not analyze how the strengthening (or reduction) of import restrictions affected the price of the middle factor (factor 3, or unskilled labor). In our analysis, the strengthening implies that $P = p_1^* - p_2^* = (-)$, and the reduction implies that $P = (+)$.

Our results suggest that it is possible for us to analyze how the trade policy change affected the middle factor in the US if we have two other pieces of information. That is, the information on the factor-intensity ranking of the middle factor (that is, which equation holds, $\theta_{31} > \theta_{32}$, or $\theta_{31} < \theta_{32}$) and the information on the position of the EWS-ratio vector, that is, the subregion P1, P2, or P3. Using these pieces of information, we can identify the Stolper-Samuelson sign pattern.

If we assume $\theta_{31} > \theta_{32}$, we know that three of the Stolper-Samuelson sign patterns hold as shown above. On the other hand, if we assume $\theta_{31} < \theta_{32}$, we can analyze similarly.

Of course, if we use econometrics, we can estimate the value of each coefficient in eq. (56), that is, the Rybczynski matrix. Therefore, we can derive the Rybczynski sign pattern and, hence, the Stolper-Samuelson sign pattern. This will be useful.

Example 2: By comparing the Cartesian coordinates of Points $R_{L2}$ and $R_{L1}$ with the EWS-ratio vector $(S', U')$, we can show some examples of a sufficient condition for a specific Stolper-Samuelson sign pattern to hold. We assume

$$(S', U') = (+, -) \leftrightarrow (S, T, U) = (+, +, -).$$

From (71), the Cartesian coordinates of Points $R_{L2}$ and $R_{L1}$ are, respectively,

$$(\frac{\theta_{K1}}{\theta_{T1}}, \frac{-\theta_{K1}}{1-\theta_{L1}} \frac{\theta_L}{\theta_K}) \; (\frac{\theta_{K2}}{\theta_{T2}}, \frac{-\theta_{K2}}{1-\theta_{L2}} \frac{\theta_L}{\theta_K}). \tag{104}$$

(i) If the EWS-ratio vector $(S', U')$ satisfies

$$\frac{\theta_{K2}}{\theta_{T2}} < S', \; \frac{-\theta_{K2}}{1-\theta_{L2}} \frac{\theta_L}{\theta_K} > U', \tag{105}$$

The EWS-ratio vector exists in the lower right of point $R_{L1}$. Hence, it exists in the subregion P1.

(ii) If the EWS-ratio vector satisfies

$$0 < \frac{\theta_{K1}}{\theta_{T1}} < S' < \frac{\theta_{K2}}{\theta_{T2}}, \text{ and } 0 > \frac{-\theta_{K1}}{1-\theta_{L1}} \frac{\theta_L}{\theta_K} > U' > \frac{-\theta_{K2}}{1-\theta_{L2}} \frac{\theta_L}{\theta_K}, \tag{106}$$

The EWS-ratio vector exists in the lower right of point $R_{L2}$ and in the upper left of point $R_{L1}$. Hence, it exists in the subregion P2.

(iii) If the EWS-ratio vector satisfies

$$0 < S' < \frac{\theta_{K1}}{\theta_{T1}}, \text{ and } 0 > U' > \frac{-\theta_{K1}}{1-\theta_{L1}} \frac{\theta_L}{\theta_K}, \tag{107}$$

The EWS-ratio vector exists in the lower right of the origin of O, and in the upper left of point $R_{L2}$. Hence, it exists in the subregion P3.

In all three cases, a strong Rybczynski result holds.

Example 3: In summary, I have shown that the position of the EWS-ratio vector determines the Rybczynski sign pattern (see eq. (91)). Notably, if extreme factors are economy-wide complements, a strong Rybczynski result holds necessarily (see Theorem 1).

Therefore, the question arises. Can we estimate the position of the EWS-ratio vector? Nakada (2016a) has shown that the EWS-ratio vector exists on the line segment. Using this relationship, he has developed a method to estimate the position of the EWS-ratio vector. That is, we can estimate it to some extent if we have the appropriate data. Nakada (2016a) derived the following results.

(i) First, he derived a sufficient condition for the EWS-ratio vector to exist in quadrant IV (that is, subregion P1, P2, or P3). In this case, extreme factors are economy-wide complements. If this holds, 'a strong Rybczynski result' holds, that is, three of the Rybczynski sign patterns hold.

(ii) Further, he derived a sufficient condition for the EWS-ratio vector to exist in a specific subregion (P1, P2, or P3). If this holds, a specific Rybczynski sign pattern holds.

In addition, Nakada (2016a) has shown that extreme factors must be economy-wide complements in some cases. In this case, it is not plausible to assume the functional form of production functions, such as Cobb-Douglas, or all-constant CES in each sector, which do not allow any two factors to be Allen-complements. Hence, we derive

$$\sigma^{ij}_h \neq (1,1,1), (c,c,c), \text{ c is constant.} \tag{108}$$

Example 4: Further, Nakada (2016b) applied Nakada's (2016a) results to data from Thailand and, in doing so, derived the factor endowment–commodity output relationship for Thailand during the period 1920-1929. He restricted the analysis to this period on account of data availability. I show the essence of his results.

In the model, Nakada (2016b) considered rice as an exportable (or commodity 1) and cotton textiles as an importable (or commodity 2). He considered land, capital, and labor as the three factors. Nakada (2016b) showed that a certain pattern of factor intensity ranking, as shown in eq. (23), holds for Thailand. Moreover, he assumed that the factor intensity ranking of the middle factor, as shown in eq. (24), holds. That is, sector 1 was relatively land intensive, sector 2 was relatively capital intensive, labor was the middle factor, and land and capital were extreme factors. He assumed that the middle factor was used relatively intensively in sector 1. He could draw the following conclusions for the data pertaining to Thailand for the period 1920-1929. The EWS-ratio vector $(S', U')$ exists in quadrant IV (or sub-regions P1-P3), in other words, capital and land, extreme factors, are economy-wide complements. Hence, a strong Rybczynski result necessarily holds.

He derived three of the Rybczynski sign patterns. However, by making a more detailed estimate, he could reduce three candidates to two.

The effects of land endowment on commodity output in sector 1 and sector 2 were positive and negative, respectively. The effects of capital endowment on commodity output in sector 1 and sector 2 were negative and positive, respectively. However, it is indeterminate how an increase in the supply of labor affected the outputs of commodities 1 and 2.

(iv) If the EWS ratio vector $(S', U')$ exists in subregion P1, the effects of labor endowment on commodity output in sector 1 and sector 2 were negative and positive, respectively.

(v) If the EWS ratio vector exists in subregion P2, the effects of labor endowment on commodity output in both sectors 1 and 2 were positive.

The results imply that Feeny's (1982, p. 28) statement that the growth in labor (or middle factor) stock was responsible for the large growth in rice output relative to textile output in Thailand might not necessarily hold.

To some extent, my results show how Chinese immigration affected commodity output in Thailand between 1920 and 1929. For example, Skinner stated, "[During 1918-1931], the Chinese flocked into Siam at an unprecedented rate...This mass influx of Chinese resulted, quite simply, from favorable conditions in Siam and unfavorable conditions in south China" (Skinner, 1957, pp. 172-174).

Example 5: Recall eqs (40), (15), and (9)

$$(S', U') = (S/T, U/T) = (g_{LK}/g_{LT}, g_{KT}/g_{LT}), \qquad (109)$$

$$g_{ih} = \Sigma_j \lambda_{ij} \varepsilon^{ij}_{h}, i, h = T, K, L, \qquad (110)$$

$$\varepsilon^{ij}_{h} = \partial \log a_{ij} / \partial \log w_h = \theta_{hj} \sigma^{ij}_{h}. \qquad (111)$$

$g_{ih}$ is the aggregate of $\varepsilon^{ij}_{h}$. Hence, we derive

$$(S', U') = (S/T, U/T) = (\frac{\Sigma_j \lambda_{Lj} \varepsilon^{Lj}_{K}}{\Sigma_j \lambda_{Lj} \varepsilon^{Lj}_{T}}, \frac{\Sigma_j \lambda_{Kj} \varepsilon^{Kj}_{T}}{\Sigma_j \lambda_{Lj} \varepsilon^{Lj}_{T}})$$

$$= \left( \frac{\Sigma_j \lambda_{Lj} \theta_{Kj} \sigma^{Lj}{}_K}{\Sigma_j \lambda_{Lj} \theta_{Tj} \sigma^{Lj}{}_T}, \frac{\Sigma_j \lambda_{Kj} \theta_{Tj} \sigma^{Kj}{}_T}{\Sigma_j \lambda_{Lj} \theta_{Tj} \sigma^{Lj}{}_T} \right). \tag{112}$$

The EWS-ratio vector contains AESs. For example, if we substitute the data on $\lambda_{ij}, \theta_{ij}$, and assume the value of AES, we can compute the Cartesian coordinates of the EWS-ratio vector. This is a type of simulation. However, as explained in Example 3, the EWS-ratio vector exists on the line-segment. This implies that the EWS-ratio vector is constrained by the data observed, hence, it cannot be arbitrary.

Moreover, by analogy with the EWS-ratio vector, I expect that the value of AES is constrained by the data observed, hence, the value cannot be arbitrary. However, I do not discuss this.

For example, if we assume the Cobb-Douglas production function in each sector, the AESs are all units:

$\sigma^{ij}{}_h = (1,1,1)$ for all *i, h, j*.

If we substitute this in eq. (112), we derive the EWS-ratio vector as follows.

$$(S', U') = (+, +).$$

It exists in quadrant I, that is, the subregions P1-P5. From eq. (92), a strong Rybczynski result holds if the EWS-ratio vector exists in the subregion P1, P2, or P3. The position of the EWS-ratio vector depends on the value of $\lambda_{ij}$ and $\theta_{ij}$. However, note that this simulation is not plausible in some cases (see (108)).

4. Conclusion

We assumed a certain pattern of factor-intensity ranking, including a certain pattern of factor-intensity ranking of the middle factor. We have assumed that sector 1 is relatively land intensive, and sector 2 is relatively capital intensive, and that labor is the middle factor, and land and capital are extreme factors. Further, we assume that the middle factor is used relatively intensively in sector 1. We analyzed the Rybczynski matrix and its sign pattern using the EWS-ratio vector and the Hadamard product. This matrix expresses the factor endowment–commodity output relationships. There are 12 patterns in total. The EWS-ratio vector boundary demarcates the boundary of the region where the EWS-ratio vector can exist. Line ij divides this region into 12 subregions. We have derived a sufficient condition for each Rybczynski sign pattern to hold. That is, the position of the EWS-ratio vector determines the Rybczynski sign pattern. A strong Rybczynski result holds for some subregions. We derived a sufficient condition for a strong Rybczynski result to hold (or not to hold) in a systematic manner. Notably, if the EWS-ratio vector $(S', U')$ exists in quadrant

IV (or subregions P1-P3); in other words, if capital and land, extreme factors, are economy-wide complements, a strong Rybczynski result holds necessarily. This result itself might not sound new. However, expressing the theorem using the EWS-ratio vector is novel. This enables us to perform further analysis. We also analyzed the Stolper-Samuelson matrix and its sign pattern, which expresses the commodity price–factor price relationships. Some applications are presented.

Can we estimate the position of the EWS-ratio vector? As I stated in Section 3, Nakada (2016a) has shown that the EWS-ratio vector exists on the line segment. Using this relationship, the author developed a method to estimate the position of the EWS-ratio vector. That is, we can estimate it to some extent if we have the appropriate data. Further, Nakada (2016b) applied Nakada's (2016a) results to data from Thailand and, in doing so, derived the factor endowment–commodity output relationship for Thailand during the period 1920 to 1929. On this, see Section 3.

This article provides the basis for such applications. It will be useful for efforts to derive the factor endowment–commodity output relationships in some countries. For example, this study contributes to international and energy economics. For example, the EWS-ratio vector is useful for the analysis of functional relations in a 3 x 2 model of another type, that is, a 3 x 2 model with three factors (capital, labor, and imported energy), for example. In this model, one of factor payments is exogenous. On this, see Nakada (2016c).

Appendix A: Derivation of important relationships among EWS

This appendix is a modified version of Nakada (2015b). Thompson (1985, p. 618) stated, 'Aggregate substitution between factors $h$ and $k$ is expressed by the substitution term

$$s_{kh} = \Sigma_j x_j \partial a_{kj} / \partial w_h \quad [, k, h = 1, 2, 3]. \tag{A1}$$

The 3 x 3 matrix of substitution terms is symmetric and negative semidefinite. A result of cost minimizing behavior is

$$\Sigma_i s_{hi} w_i = 0, \quad \text{for every factor } h \quad [, h = 1, 2, 3].' \tag{A2}$$

Thompson's (1985) definition of these symbols is similar to the definitions in this article, but his explanation seems too short. The cost minimizing behavior implies that each $a_{ij}$ function is homogeneous of degree zero for all input prices (see eq. (3), note 5). From this, we can derive Thompson's (1985) result (A2). We prove it below.

Recall eq. (9),

$$\varepsilon^{ij}_h = \partial \log a_{ij} / \partial \log w_h = \theta_{hj} \sigma^{ij}_h. \tag{A3}$$

From eq. (A3), we obtain

$$\partial a_{ij} / \partial w_h = \varepsilon^{ij}{}_h a_{ij} / w_h, \ i, h = T, K, L, j = 1, 2. \tag{A4}$$

Replacing $s_{kh}$ in (A1) with $s_{ih}$, we derive

$$s_{ih} = \Sigma_j x_j \partial a_{ij} / \partial w_h, \ i, h = T, K, L. \tag{A5}$$

Substituting (A4) in (A5), we obtainjjj

$$s_{ih} = \Sigma_j x_j \varepsilon^{ij}{}_h a_{ij} / w_h, i, h = T, K, L. \tag{A6}$$

Because each $a_{ij}$ function is homogeneous of degree zero (recall eq. (12)):

$$\Sigma_h \varepsilon^{ij}{}_h = \Sigma_h \theta_{hj} \sigma^{ij}{}_h = 0, \ i = T, K, L, j = 1, 2. \tag{A7}$$

From eqs (A6) and (A7), we can show that

$$\Sigma_h s_{ih} w_h = 0, \ i = T, K, L. \tag{A8}$$

This is equivalent to eq. (A2).

AESs are symmetric in the sense that (see eq. (10))

$$\sigma^{ij}{}_h = \sigma^{hj}{}_i. \tag{A9}$$

Additionally, according to BC (p. 33), 'Given the assumption that production functions are strictly quasi-concave and linearly homogeneous,' (see eq. (11))

$$\sigma^{ij}{}_i < 0. \tag{A10}$$

From eqs (A6), (A3), and (A9), we can show that

$$s_{ih} = s_{hi}, \tag{A11}$$

specifically, aggregate substitutions are symmetric. Substitute eq. (A10) in eq. (A3) to derive $\varepsilon^{ij}{}_i < 0$. By substituting this equation in eq. (A6), we obtain

$$s_{ii} < 0. \tag{A12}$$

Next, we analyze $s_{LL}$ in a similar way as that used by BC (p. 33) in analyzing AES ($\sigma^{Lj}{}_L$). Eliminating $s_{TL}$ and $s_{KL}$ from eq. (A8), we derive

$$s_{LL} = \frac{1}{(w_L)^2} \{ w_T (w_T s_{TT} + w_K s_{TK}) + w_K (w_T s_{KT} + w_K s_{KK}) \}. \tag{A13}$$

Transform (A13):

$$s_{LL} = \mathbf{x} \cdot \mathbf{A}\mathbf{x}, \tag{A14}$$

where $\mathbf{x}$ is a vector, $\mathbf{A}$ is a matrix, and $\mathbf{x} \cdot \mathbf{A}\mathbf{x}$ is the inner product of vectors;

$$\mathbf{x} = \begin{bmatrix} w_K \\ w_T \end{bmatrix}, \mathbf{A} = \begin{bmatrix} s_{KK} & s_{KT} \\ s_{TK} & s_{TT} \end{bmatrix}.$$

To quote a passage from BC (p. 33): 'the quadratic form on the right-hand side of the expression above must be negative definite. This, in turn, implies that' [17]

$$|A| = s_{KK} s_{TT} - s_{KT} s_{TK} > 0, \tag{A15}$$

where $|A|$ is the 2 x 2 determinant. Transform eq. (A6) to derive

$$s_{ih} = \Sigma_j \lambda_{ij} \varepsilon^{ij}_h V_i / w_h = g_{ih} V_i / w_h, \quad i, h = T, K, L. \tag{A16}$$

This equation shows how aggregate substitution and EWS are related.[18] From eq. (A16), $g_{ih}$ is not symmetric. Specifically, $g_{ih} \neq g_{hi}$, $i \neq h$ in general.

Substituting eq. (A16) in eq. (A15), we obtain

$$g_{KK} g_{TT} - g_{KT} g_{TK} > 0. \tag{A17}$$

This equation shows that JE's proof is impossible. Next, we show the disproof of JE. JE (p. 75) define $\sigma_i^k, i, k = 1, 2, 3$, as EWS. In subsection 5.2.4 (p. 90), JE states, 'First we assume that the two extreme factors [factors 1 and 2] are perfect complements in the sense that any factor price change does not alter the ratio of the intensities of their use ($\sigma_1^k = \sigma_2^k [, k = 1, 2, 3]$).'

Here, for the authors, 'perfect complementarity' implies $\sigma_1^k = \sigma_2^k$. If we replace $\sigma_i^k$ with $g_{ih}$, this implies that

$$g_{Th} = g_{Kh}, \, h = T, K, L \leftrightarrow g_{TT} = g_{KT}, \, g_{TK} = g_{KK}, \, g_{TL} = g_{KL}. \tag{A18}$$

In other words, the authors found that the set of three equations holds for EWS under the assumption of 'perfect complementarity.' Next, the authors used this set to prove how commodity prices affect factor prices.

If we compare eq. (A18) with eq. (A17), we find that the latter is not consistent with the former. That is, if eq. (A18) holds, L.H.S. of (A17) equals zero. Hence, JE's result is impossible. Specifically, the authors fails to explain what 'perfect complementarity' implies. In summary, their proof is not plausible.

In subsection 5.2.5 (p. 91), JE's analysis is similarly to subsection 5.2.4. The authors assume that extreme factor (factor 2) is a perfect complement of the middle factor (factor 3). The authors state that they derive $\sigma_3^1 = \sigma_2^1$. In the authors' context, this implies $\sigma_3^k = \sigma_2^k$, $k = 1, 2, 3$. We can prove in a similar fashion that this is impossible.

Appendix B:

$\Delta$ is the determinant of matrix **A**, the coefficient matrix of a system of linear equations (see eq. (46)). We can show that $\Delta < 0$. $\Delta$ is equivalent to the 3 x 3 determinant $D$ in BC, and it was proved that $D < 0$ (see BC (p. 25-26)). However, BC's method requires some technique. We show the proof using the important relationship among EWSs. From eqs (46) and (21), we derive

$$\Delta = \det(\mathbf{A}) = \begin{vmatrix} \theta_{T1} & \theta_{K1} & \theta_{L1} & 0 & 0 \\ \theta_{T2} & \theta_{K2} & \theta_{L2} & 0 & 0 \\ g_{TT} & g_{TK} & g_{TL} & \lambda_{T1} & \lambda_{T2} \\ g_{KT} & g_{KK} & g_{KL} & \lambda_{K1} & \lambda_{K2} \\ g_{LT} & g_{LK} & g_{LL} & \lambda_{L1} & \lambda_{L2} \end{vmatrix}. \tag{B1}$$

Sum columns 1 and 2 in column 3, and subtract row 2 from row 1. We have

$$\Delta = \begin{vmatrix} A & B & 0 & 0 & 0 \\ \theta_{T2} & \theta_{K2} & 1 & 0 & 0 \\ g_{TT} & g_{TK} & 0 & \lambda_{T1} & \lambda_{T2} \\ g_{KT} & g_{KK} & 0 & \lambda_{K1} & \lambda_{K2} \\ g_{LT} & g_{LK} & 0 & \lambda_{L1} & \lambda_{L2} \end{vmatrix}, \tag{B2}$$

where we may recall eq. (26), that is, $(A, B, E) = (\theta_{T1} - \theta_{T2}, \theta_{K1} - \theta_{K2}, \theta_{L1} - \theta_{L2})$. Express the above as a cofactor expansion along the third column:

$$\Delta = 1(-1)^{2+3} \begin{vmatrix} A & B & 0 & 0 \\ g_{TT} & g_{TK} & \lambda_{T1} & \lambda_{T2} \\ g_{KT} & g_{KK} & \lambda_{K1} & \lambda_{K2} \\ g_{LT} & g_{LK} & \lambda_{L1} & \lambda_{L2} \end{vmatrix}. \tag{B3}$$

Recall eq. (17), that is, $g_{ih} = (\theta_h / \theta_i) g_{hi}$, $i, h = T, K, L$, and $\lambda_{ij} = (\theta_j / \theta_i) \theta_{ij}$. Using these equations, transform eq. (B3):

$$\Delta = -\begin{vmatrix} A & B & 0 & 0 \\ g_{TT}\theta_T/\theta_T & g_{KT}\theta_K/\theta_T & \theta_{T1}\theta_1/\theta_T & \theta_{T2}\theta_2/\theta_T \\ g_{TK}\theta_T/\theta_K & g_{KK}\theta_K/\theta_K & \theta_{K1}\theta_1/\theta_K & \theta_{K2}\theta_2/\theta_K \\ g_{TL}\theta_T/\theta_L & g_{KL}\theta_K/\theta_L & \theta_{L1}\theta_1/\theta_L & \theta_{L2}\theta_2/\theta_L \end{vmatrix}.$$ (B4)

Divide rows 2, 3, and 4 by $1/\theta_T, 1/\theta_K,$ and $1/\theta_L$, respectively, and divide columns 3 and 4 by $\theta_1$ and $\theta_2$, respectively, to derive:

$$\Delta = -\begin{vmatrix} A & B & 0 & 0 \\ g_{TT}\theta_T & g_{KT}\theta_K & \theta_{T1} & \theta_{T2} \\ g_{TK}\theta_T & g_{KK}\theta_K & \theta_{K1} & \theta_{K2} \\ g_{TL}\theta_T & g_{KL}\theta_K & \theta_{L1} & \theta_{L2} \end{vmatrix}\theta',$$ (B5)

where $\theta' = \theta_1\theta_2/\theta_T\theta_K\theta_L$. Sum rows 2 and 3 in column 4, and subtract columns 4 from row 3. We have

$$\Delta = -\begin{vmatrix} A & B & 0 & 0 \\ g_{TT}\theta_T & g_{KT}\theta_K & A & \theta_{T2} \\ g_{TK}\theta_T & g_{KK}\theta_K & B & \theta_{K2} \\ 0 & 0 & 0 & 1 \end{vmatrix}\theta'.$$ (B6)

Express the above as a cofactor expansion along the fourth row, and permutate rows 2 and 3. We have

$$\Delta = -(1)(-1)^{4+4}(-1)\begin{vmatrix} A & B & 0 \\ g_{TK}\theta_T & g_{KK}\theta_K & B \\ g_{TT}\theta_T & g_{KT}\theta_K & A \end{vmatrix}\theta'.$$ (B7)

From eq. (17), we derive $g_{TK}\theta_T = g_{KT}\theta_K$. Using this equation, expand eq. (B7) to derive

$$\Delta = \theta'[A^2 g_{KK}\theta_K + B^2 g_{TT}\theta_T - 2AB g_{KT}\theta_K].$$ (B8)

Transform eq. (B8) to derive:

$$\Delta = \theta'[g_{KK}\theta_K + x^2 g_{TT}\theta_T - 2x g_{KT}\theta_K]A^2,$$ (B9)

where $x = B/A$. This is a quadratic formula. From eq. (18), we have

$$g_{TT}\theta_T < 0.$$ (B10)

Hence, the coefficient of $x^2$ in eq. (B9) is negative. The quarter of discriminant of eq. (B9) is

$$D/4 = (g_{KT}\theta_K)^2 - g_{KK}\theta_K g_{TT}\theta_T.$$ (B11)

From eq. (17), we derive $g_{KT}\theta_K = g_{TK}\theta_T$. Substitute this equation in eq. (B11) to derive:

$$D/4 = -[g_{KK}g_{TT} - g_{KT}g_{TK}]\theta_K\theta_T.$$ (B12)

Recall eq. (A17), $g_{KK}g_{TT} - g_{KT}g_{TK} > 0.$ Substitute this in eq. (B12) to derive:

$$D/4 < 0. \tag{B13}$$

From eqs (B10) and (B13), we have

$$\Delta < 0. \tag{B14}$$

Using eqs (19) and (17), transform eq. (B8) to derive:

$$\Delta = -\theta'[(B+A)^2 g_{KT}\theta_K + g_{LK}\theta_L A^2 + g_{LT}\theta_L B^2]. \tag{B15}$$

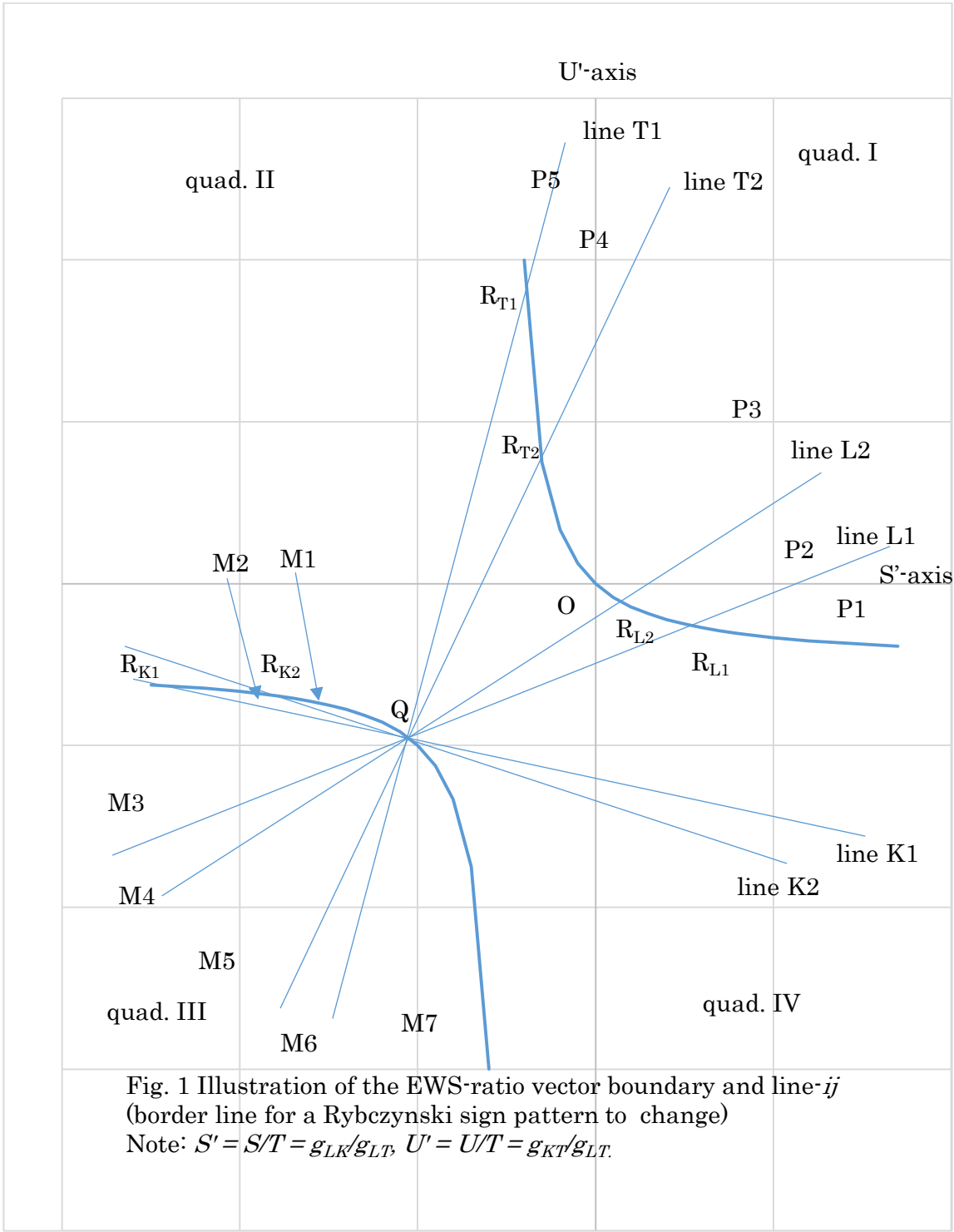

Fig. 1 Illustration of the EWS-ratio vector boundary and line-$ij$
(border line for a Rybczynski sign pattern to change)
Note: $S' = S/T = g_{LK}/g_{LT}$, $U' = U/T = g_{KT}/g_{LT.}$

---

[1] BC have derived a conclusion (p. 34) that an increase in the supply of one factor, at constant commodity prices, will increase the output of the commodity using the expanding factor relatively intensively and reduce the output of the other commodity.

[2] The authors did not show these results using the sign pattern as shown in this article, but using the ranking form such as $w_T^* > p_1^* > w_L^*(=0) > p_2^* > w_K^*$ if we use our expression. On this, see JE (p. 79, eq. (22)), for

example. However, their technique requires some skill and is not easy. Additionally, it is not useful for the analysis of a sufficient condition for a Rybczynski sign pattern to hold. The analysis based on computation is far easier. Additionally, Thompson (1993) also used a diagrammatic technique developed by JE, and supplemented JE's analysis. He derived 11 patterns of ranking form in total. Apparently, for some cases, two ranking forms correspond to the same sign pattern.

[3] Suzuki (1983) assumed that capital and land (middle factor and extreme factor, respectively) were 'perfect complements' in each sector, and derived the implications using AES, that is, '$\sigma^j_{KK} = \sigma^j_{KT} < 0, \sigma^j_{LK} = \sigma^j_{LT} > 0$, and $\sigma^j_{KT} = \sigma^j_{TT} < 0.$' $\sigma^j_{ik}$ is the AES between the ith and the kth factors in the jth industry. Suzuki used this in his disproof. BC (p33) derived the relationship for AES on the assumption that the production functions were strictly quasi-concave and linearly homogeneous, i.e.: $\sigma^j_{KK}\sigma^j_{TT} - (\sigma^j_{KT})^2 > 0.$ If we compare this inequality with Suzuki's equations, we find that the latter is not consistent with the former. On this, see Nakada (2015a).

[4] In subsections 5.2.1 to 5.2.5 (p. 86-92), JE analyzed the cases shown below. That is, (1) factor intensity of the middle factor is the same, or $\theta_{L1} = \theta_{L2}$ in our expression, (2) extreme factors are independent, or $g_{TK} = 0$ in our expression, (3) all factors including extreme factors are substitutes, (4) extreme factors are perfect complements, (5) the middle factor and an extreme factor are perfect complements. Specifically, both (1) and (2) are special cases. In the case of (3), JE (p. 88) assumed that 'the middle factor is used more intensively in $x_1$ [or sector 1] than in $x_2$ [or sector 2],' that is, $\theta_{L1} > \theta_{L2}$ in our expression. JE only showed two patterns of the commodity price–factor price relationships that hold. The explanations in (3) are complicated. I am uncertain whether they are plausible. If all factors are substitutes, $(S, T, U) = (+, +, +)$ holds, hence, $(S', U') = (+, +)$ holds (see eq. (43)). Therefore, as we show in section 3 in this article, the EWS-ratio vector exists in quadrant I, that is, in the subregion P1 to P5. This implies that five patterns of the commodity price–factor price relationship hold. This is not discussed further.

[5] From eq. (A16), if factors *i* and *h* are aggregate complements, they are economy-wide complements, and vice versa. Takayama (1982) showed only one sufficient condition for a strong Rybczynski result to hold. Suzuki (1987) derived a similar result. In Suzuki (1987, Chapter 1, p. 17-26), the author assumed that extreme factors are 'Allen-complements' in each sector (p. 23), and he derived a strong Rybczynski result. Apparently, if extreme factors are

Allen-complements in each sector, extreme factors are aggregate complements, but not vice versa.

[6] Additionally, Ban (2010) modified an important basic assumption. She assumed that commodity prices are endogenous. She analyzed how factor endowments affected factor prices in a theoretical study.

[7] Additionally, as I showed in Nakada (2016a), in some cases, it is not plausible to assume that production functions are of a Cobb-Douglas type, or an all-constant CES type in each sector, which do not allow any two factors to be Allen-complements, as Thompson (1995) assumed. Moreover, it is not plausible to assume that production functions are of the two-level CES type, as Ban (2007a) assumed.

[8] EWS contains AES in two sectors. Strangely, JE did not mention AES at all. There are nine EWSs. I show that only three EWSs are needed for the analysis. Absolute value of EWS is not important to analyze a sufficient condition for each Rybczynski sign pattern to hold. Only by defining the EWS-ratio vector can we analyze it systematically using the figure in two dimensions.

[9] For example, Nakada (2016b) applied Nakada's (2016a) results to data from Thailand and, in doing so, derived the factor endowment–commodity output relationship for Thailand during the period 1920 -1929. To some extent, these results show how Chinese immigration affected commodity output in Thailand between 1920 and 1929.

[10] Our method using a 5 x 5 matrix does not require special techniques, which other studies used. For example, BC transformed some equations using some techniques and made a system of linear equations using a 3 x 3 matrix. On the other hand, in section 3 (p. 73-77), JE used other techniques, and made a system of three linear equations. In fact, these methods are not easy to reapply.

[11] We assume that sector 1 is relatively land intensive, and sector 2 is relatively capital intensive, and that labor is the middle factor, and land and capital are extreme factors. Further, we assume that the middle factor is used relatively intensively in sector 1.

[12] However, estimating the values of parameters belongs to a partial equilibrium analysis.

[13] In section 4, Teramachi (2015, p. 50) showed 12 patterns of '$J$ sign patterns', which express the commodity price–factor price relationships ($J \equiv \partial \log W / \partial \log P$, $J = w_i{}^*/p_j{}^*$ in our expression). According to Teramachi (2015, p. 52), this expression does not show a one-to-one correspondence with the ranking form of JE. That is, one $J$ sign pattern can include two ranking forms in JE. In section 5 (p. 55-61), Teramachi (2015) showed some sufficient conditions for $J$ sign pattern to hold. He analyzed the cases shown below (Case A-F). That is, (A) specific factors model, (B) extreme factors are independent, (C) extreme factors are complements (or

perfect complements), (D) factor intensity of the middle factor is the same, (E) the middle factor and an extreme factor are perfect complements, (F) all factors are substitutes. His analysis is mainly based on the condition that JE showed, as he stated. That is, the sufficient conditions that Teramachi showed are similar to those that JE analyzed. Out of the six sufficient conditions that Teramachi showed, five conditions do not show a one-to-one correspondence with $J$ sign pattern. See the table in Teramachi (2015, p. 61).

[14] For example, Ban (2008) showed the factor-intensity ranking as follows. That is, $\theta_{S1}/\theta_{S2} > 1 > \theta_{L1}/\theta_{L2} > \theta_{K1}/\theta_{K2}$, where $\theta_{ij}$ denotes the cost share (distributive share in our expression); $S$ is the skilled labor, $K$ capital, and $L$ unskilled labor. This implies that unskilled labor is the middle factor, and skilled labor and capital are extreme factors.

[15] Ban (2008, Appendix table) did not compute the distributive share, based on which we show the factor-intensity ranking for the middle factor, that is, whether $\theta_{L1} > \theta_{L2}$ or $\theta_{L1} < \theta_{L2}$ holds, for example, if unskilled labor ($L$) is the middle factor. She only showed whether $a_{L1} > a_{L2}$ or $a_{L1} < a_{L2}$ held, if we use our expression. Similarly, Ban (2011, chapter 4, p. 107, Appendix Table 4-1) did not compute the distributive share. This is confusing.

[16] Some explanation is required. Samuelson (1953, Chapter 4, p. 59) defines the function, $v_i = f^i(x, w_1, \cdots, w_n), (i = 1, \cdots, n)$. $v_i$ is 'an optimum value for each productive factor' to derive 'the minimum total cost for each output (p. 58),' $x$ is production, and $w_i$ is 'prices of productive factors.' Samuelson (1953, Chapter 4, p. 68) stated that $v_i$ 'must be homogeneous of order zero in the variables $(w_1, \cdots, w_n)$, $x$ being constant' (see also Samuelson (1986, chapter 4, eq. (5) in p. 61; eq. (52) in p. 70)). This implies that from the condition of cost minimization, we can show that $a_{ij}$ is homogeneous of degree zero in all input prices.

[17] Takayama (1982, p. 5, Theorem 1, note 5) analyzed the general $m$ x $n$ model, and he stated that because 'substitution matrix' **S** is negative semidefinite and $R(\mathbf{S}) = m - 1$, the $(m-1)$ x $(m-1)$ matrix is negative definite, from which $s_{ii} < 0$, $i = 1, 2, \ldots, m$. $R(\mathbf{S})$ denotes the rank of a particular matrix, and $\mathbf{S} = [s_{ih}]$.

[18] Teramachi (1993, p. 44) showed the equation equivalent to (A16).